\documentclass[fleqn,usenatbib]{mnras}

\usepackage{newtxtext,newtxmath}

\usepackage[T1]{fontenc}

\DeclareRobustCommand{\VAN}[3]{#2}
\let\VANthebibliography\thebibliography
\def\thebibliography{\DeclareRobustCommand{\VAN}[3]{##3}\VANthebibliography}

\usepackage{graphicx}
\usepackage{amsmath}
\usepackage{url}
\usepackage{hyperref}

\title[J01020100-7122208: an accreted star]{ J01020100-7122208: an accreted evolved blue straggler that wasn't ejected from a supermassive black hole}

\author[D. Brito-Silva et al.]{
Danielle de Brito Silva,$^{1}$\thanks{E-mail: danielle.debrito@mail.udp.cl}
Paula Jofr\'e,$^{1}$
Douglas Bourbert,$^{2}$
Sergey E. Koposov,$^{3,4}$
Jose L. Prieto$^{1,5}$
\newauthor and Keith Hawkins$^{6}$
\\
$^{1}$N\'ucleo de Astronom\'ia, Universidad Diego Portales, Ej\'ercito 441, Santiago, Chile\\
$^{2}$ Magdalen College, University of Oxford, High Street, Oxford OX1 4AU, UK\\
$^{3}$Institute for Astronomy, University of Edinburgh, Royal Observatory, Blackford Hill, Edinburgh EH9 3HJ, UK\\
$^{4}$Institute of Astronomy, University of Cambridge, Madingley Road, Cambridge CB3 0HA, UK\\
$^{5}$Millennium Institute of Astrophysics, Santiago, Chile \\
$^{6}$Department of Astronomy, The University of Texas at Austin, 2515 Speedway Boulevard, Austin, TX 78712, USA
}

\date{Accepted XXX. Received YYY; in original form ZZZ}

\pubyear{2021}

\begin{document}
\label{firstpage}
\pagerange{\pageref{firstpage}--\pageref{lastpage}}
\maketitle

\begin{abstract}
J01020100-7122208 is a star whose origin and nature still challenges us.
It was first believed to be a yellow super giant ejected from the Small Magellanic Cloud, but it was more recently claimed to be a red giant accelerated by the Milky Way's central black hole. 
In order to unveil its nature, we analysed photometric, astrometric and high resolution spectroscopic observations to estimate the orbit, age, and 16 elemental abundances. Our results show that this star has a retrograde and highly-eccentric orbit, $e=0.914_{-0.020}^{+0.016}$. Correspondingly, it likely crossed the Galactic disk at $550\;\mathrm{pc}$ from the Galactic centre. We obtained a spectroscopic mass and age of $1.09\pm0.10$ $M_\odot$ and  $4.51\pm1.44$ Gyr respectively. Its chemical composition is similar to the abundance of other retrograde halo stars. We found that the star is enriched in europium, having [Eu/Fe] = 0.93 $\pm$ 0.24, and is more metal-poor than reported in the literature, with [Fe/H] = -1.30 $\pm$ 0.10. This information was used to conclude that  J01020100-7122208 is likely not a star ejected from the central black of the Milky Way or from the Small Magellanic Cloud. Instead, we propose that it is simply a halo star which was likely accreted by the Milky Way in the distant past  but its mass and age suggest it is probably an evolved blue straggler. 
\end{abstract}


\section{Introduction} \label{sec:intro}

\textit{Gaia} \citep{GaiaCollaboration+2016b} is revolutionising our understanding of the Milky Way, providing astrometric information not only for our Galaxy's stars but also for stars from satellite galaxies. With the \textit{Gaia} Early Data Release 3 (\textit{Gaia}  EDR3; \citealt{brown2021gaia}), it is possible to describe the detailed orbits of many stars, including J01020100-7122208. J01020100-7122208 is a star that has a velocity of 300 km s$^{-1}$ whose origin has been analysed in recent years \citep{neugent2018runaway,massey2018runaway}, yet no consensus has been reached. Understanding the nature of high-velocity stars is important because from them we can learn, for example, about how frequently encounters between stars and the central black hole of the Milky Way are (e.g. \citealt{rossi2017joint}), as well as what is the escape velocity of the Galaxy as a function of Galactocentric radius (e.g. \citealt{piffl2014rave}). Therefore, reaching a consensus of J01020100-7122208 is important. 

This star was a serendipitous discovery. \cite{neugent2010yellow} carried out a study with the objective of identifying yellow stars in the direction of the Small Magellanic Cloud (SMC) based on the radial velocities (RVs) of the objects in that field. Their motivation  was that by studying yellow stars it is possible to test stellar evolutionary theory which can help us to interpret the light of distant galaxies. The authors observed 496 stars using the multi-object spectrometer Hydra at Cerro Tololo 4 m telescope. They determined RVs by cross-correlating their spectra using the Ca II triplet. 
They then compared their results for individual stars with the mean RV of the SMC 
and considered that stars with RVs similar to the value of that dwarf galaxy (160 km s$^{-1}$) were candidate members. In cases where the RV results obtained using this method were inconclusive, they complemented their analysis by using the luminosity-sensitive line O I $\lambda$ 7774: stars with measurable amounts of O I $\lambda$ 7774 should be supergiants (making it possible that the objects were in the SMC) while stars without measurable  O I $\lambda$ 7774 should be foreground stars from the Milky Way. With this methodology, \cite{neugent2010yellow} found 176 stars that were candidate members of the SMC. They found one star with an extreme heliocentric RV of about 300 km s$^{-1}$: J01020100-7122208. 

Several years later, \citet[][hereafter Neu18]{neugent2018runaway} presented a more detailed spectroscopic analysis of J01020100-7122208, using one spectrum from the spectrometer Hydra at Cerro Tololo 4m telescope, one spectrum from Echelle on the du Pont 2.1 m telescope at Las Campanas Observatory and two spectra from MagE on the Las Campanas Baade 6.5m Magellan telescope. The spectra used have resolution ranging from 3000 to 45000.

In that study, the authors classified the star as a G5-8, with an effective temperature ($T_\mathrm{eff}$) of 4700 $\pm$ 270 K, a surface gravity ($\log g$) of 0.8, a mass of 9 $M_\odot$ and an age of 30 Myr. They determined the spectral type by comparing a spectrum of J01020100-7122208 with the spectra of different spectral standards, considering the strength of metal lines and the weakness of hydrogen lines. To calculate $T_\mathrm{eff}$ they compared de-reddened broad-band colors to model atmospheres. Finally they used Geneva evolution models to determine the mass and, as a consequence, the age of the star. With this information they concluded that J01020100-7122208 was a yellow supergiant which was likely ejected from the SMC due to a binary companion that exploded as a supernova. This would also explain its high velocity.

As soon as \textit{Gaia}  DR2 \citep{brown2018gaia} was released, J01020100-7122208 was studied again by \citet[][hereafter Mas18]{massey2018runaway}. The authors noticed that the star's parallax was too large to be consistent with an origin in the SMC.  They thus claimed that J01020100-7122208  was instead in the Galactic halo. They performed a new spectral analysis using the same spectra used in Neu18,
estimating a $T_\mathrm{eff}$ of 4800 $\pm$ 100 K, a $\log g$ of 2.2, an age of 180 Myr,  a mass of 3-4 $M_\odot$ and a metallicity of $\mathrm{[Fe/H]} = -0.5$. They compared these values with evolutionary tracks and concluded that the star was in the giant or early asymptotic giant branch stage, i.e. it was not a yellow supergiant. From the  age and metallicity they further concluded that the star was born in the Galactic disk. After extensively discussing several different scenarios to explain the radial velocity, Mas18 concluded that J01020100-7122208 was likely ejected from the central black hole of the Milky Way. The basis for this argument was that, when Mas18 integrated the orbit of the star back in time, they found that the star passed close to the centre of the Galaxy. The authors commented that later \textit{Gaia} data releases could alter this conclusion.

Now is an interesting moment to revisit the origin of this object with a new data release from \textit{Gaia} \citep{brown2021gaia}. Furthermore,  methodologies to determine ages have improved, thanks to the improved astrometry from \textit{Gaia}  and information from spectroscopy. Last but not least,  high resolution spectra can be used to determine chemical abundances in addition to stellar parameters. The later has proven very useful to study the origins of stars in the Galaxy \citep{nissen2010two,hawkins2018fastest,matsuno2019origin,kordopatis2020chemodynamics,das2020ages}, because different chemical elements are produced  by different nucleosynthetic channels. By understanding the connection of these channels with the star formation timescales of the Galaxy and its satellites, one can shed light on the formation history of certain stars.
Other works in the literature such as \cite{hansen2016chemical} and \cite{hawkins2018fastest} have already used chemical abundances to explore the origin of high velocity stars in the Galaxy, proving the power of using this information about stars to unveil their origins.

In this paper, we perform a detailed analysis of J01020100-7122208 with the goal of shedding light on its origin. To do so, we consider the latest \textit{Gaia}  data (\textit{Gaia}  EDR3; \citealt{brown2021gaia}) and obtain high resolution spectra to determine stellar parameters and chemical abundances of 16 elements, ages and orbital parameters. It is the first time chemical abundances are used to study this star. In Section \ref{sec:data} we present the data used in our work. In Section \ref{sec:methods}, we describe the analysis to calculate the orbits, age, stellar parameters and chemical abundances of J01020100-7122208, and in Section \ref{sec:results} we present our results. In Section \ref{sec:discussion}, we discuss our findings. Finally, in Section \ref{sec:conclusion} we present our conclusions.

\section{Data}\label{sec:data}
The astrometric and photometric information of J01020100-7122208 was taken from \textit{Gaia}  EDR3 and can be found in Table \ref{tab:gaia_info}. 

\begin{figure*}
\centering
\includegraphics[width=19cm]{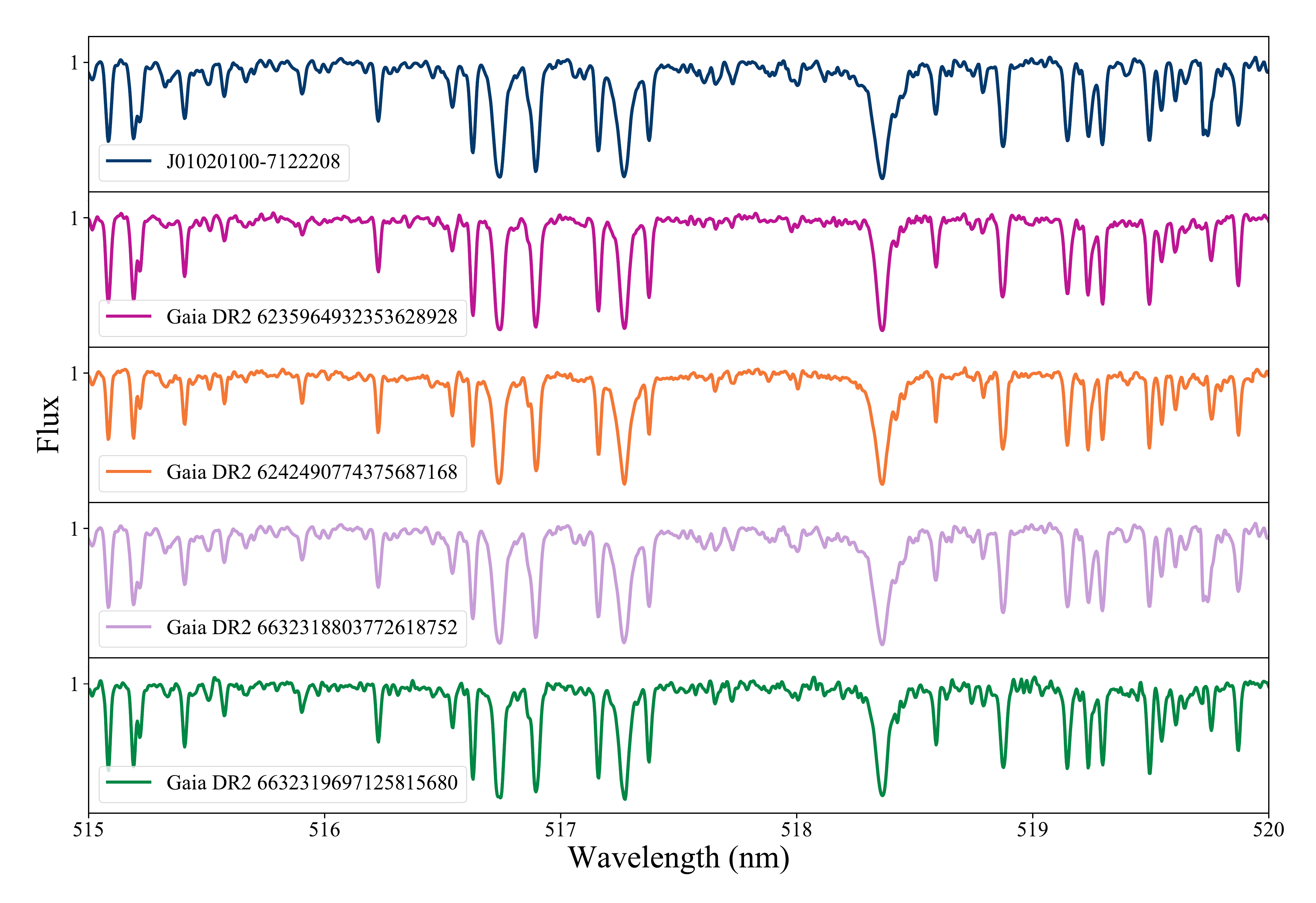}
\caption{Examples of spectra. The blue spectrum at the top corresponds to J01020100-7122208 while the magenta, orange, lilac and green ones correspond to spectra of stars used to validate our spectral results.}
\label{fig:spectrum}
\end{figure*}

\begin{table}
\caption{Properties of J01020100-7122208. Radial velocity not available in \textit{Gaia}  EDR3.} 
\label{tab:gaia_info} 
\centering
\begin{tabular}{c c}
\hline\hline

Property & Information  \\ \hline
2MASS ID & J01020100-7122208  \\
Gaia ID & 4690790008835586304  \\      
R.A. (deg) & 15.504 $\pm$ 0.011 \\
Dec. (deg) & -71.373 $\pm$ 0.010 \\
Parallax (mas) & 0.105 $\pm$ 0.012\\
Proper motion - R.A. (mas/yr) & 8.630 $\pm$ 0.017 \\
Proper motion - Dec. (mas/yr) & -0.938 $\pm$ 0.013 \\
G (mag) & 13.36 \\
\hline
\end{tabular}
\end{table}

We used two high-resolution spectra taken using the \textsc{MIKE} spectrograph on the Clay Telescope at Las Campanas Observatory and reduced using CarPy \citep{kelson2003optimal}, the standard pipeline for data reduction with that instrument. The first high-resolution spectrum (R $\approx$ 25 000) was observed on (UT) 2013 January 7 and has a signal-to-noise ratio (S/N) of about 100 pixel$^{-1}$. It allowed us to determine stellar parameters, chemical abundances and RV. The second spectrum was taken on (UT) 2019 August 27 and has a greater resolution of (R $\approx$ 55 000) but a S/N of only 40 pixel$^{-1}$. The second spectrum was solely used to obtain a second RV measurement and so to rule out a possible binary nature. Any variation of RV across several years  might indicate that this star is part of a binary system, affecting the determination of mass and age.

In addition, we considered a sample of MIKE spectra of 7 metal-poor halo stars. 
These spectra have typical S/N of 100 pixel$^{-1}$ and resolution of about 40 000. The stars were used as a control sample to validate our spectral results. These stars have been selected from \textsc{APOGEE} DR14 \citep{Abolfathi2018,holtzman2018apogee} and have been analysed by Carrillo et al, (submitted, hereafter Car21). The control sample stars satisfy 4300 K $\leq$ $\mathrm{T_{eff}}$ $\leq$ 5000 K, 0.9 $\leq$ $\log g$ $\leq$ 2.7 and -1.75 $\leq$ [Fe/H] $\leq$ -1.0.
It is important for us to validate the stellar parameters of J01020100-7122208, since previous works of Neu18 and Mas18 do not agree on them.

In Figure~\ref{fig:spectrum} we display an example region of the normalised and RV corrected spectrum of J01020100-7122208 (in blue at the top) along with 4 stars (Gaia DR2 6235964932353628928, Gaia DR2 6242490774375687168, Gaia DR2 6632318803772618752, Gaia DR2 6632319697125815680) from our control sample. 

\section{Analysis}\label{sec:methods}

\subsection{Spectroscopy}

We determined radial velocities, stellar parameters and chemical abundances using a pipeline developed by us based on the code for spectral analysis {\tt iSpec} \citep{Blanco2014,Blanco2019} and \textsc{IRAF} tasks \citep{iraf1993} were used to merge the orders of the 2D reduced spectra.

We first normalized the spectra of J01020100-7122208 and of the control sample stars order-by-order using  3-degree splines every 5 nm. We also performed the RV correction order-by-order, by cross-correlating the observed spectra with a line mask from a spectrum of Arcturus from Atlas (provided with {\tt iSpec}). We did it order-by-order to avoid problems related to the wavelength calibration of our data. At this step we obtained RVs for each order. We adopted as the star's RV the mean of the values calculated for each order and as the uncertainty the standard deviation.
We visually inspected the spectra, guaranteeing that the absorption lines were aligned with the lines in the laboratory rest frame.
From the spectrum acquired in 2013 we obtained a RV of 296.27 $\pm$ 0.17 km s$^{-1}$ while from the spectrum acquired in 2019 the calculated RV value is 296.24 $\pm$ 0.25 km s$^{-1}$. Both values agree within 1 sigma, therefore we find no evidence that this star is currently in a binary system.

\begin{figure}
\centering
\includegraphics[width=9cm]{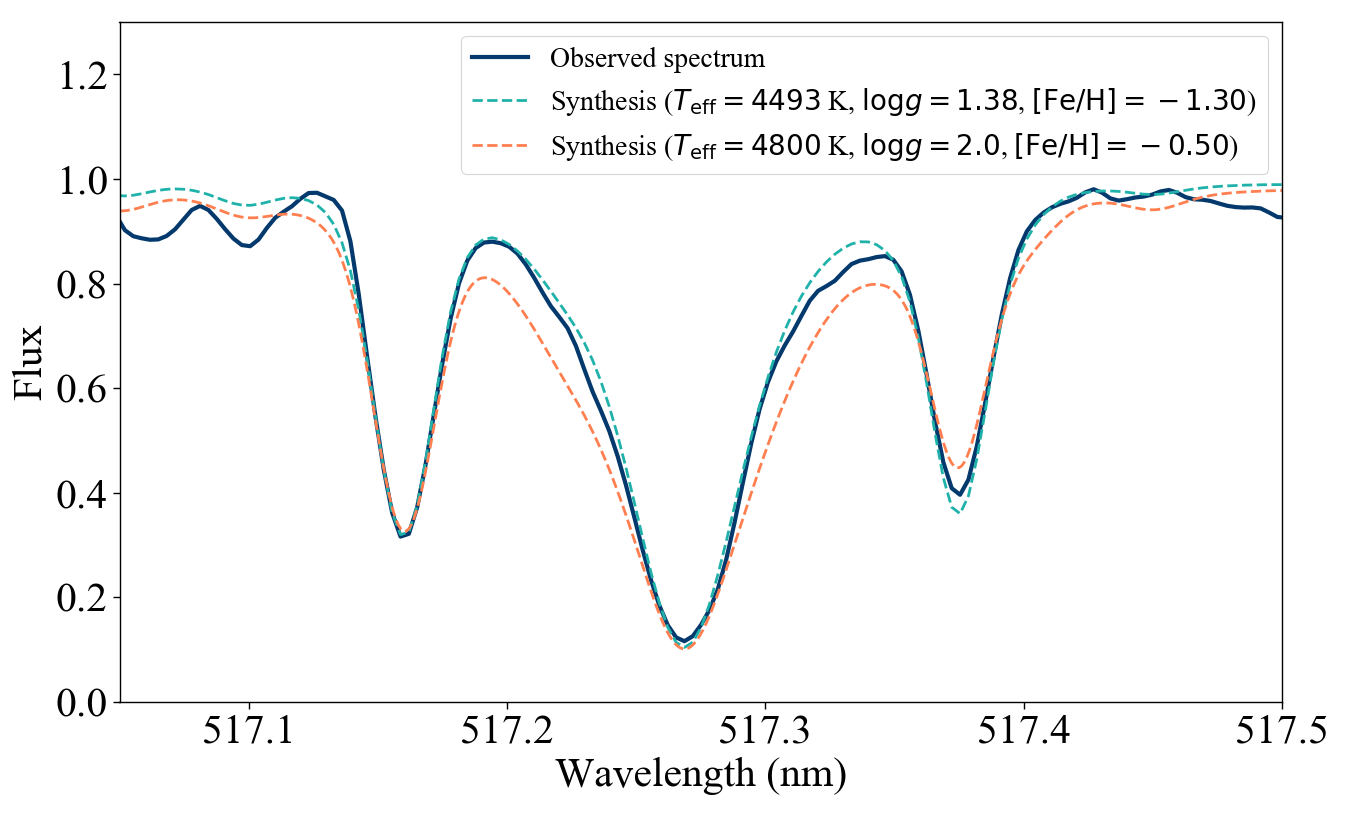}
\vspace{-1.5cm}
\includegraphics[width=9cm]{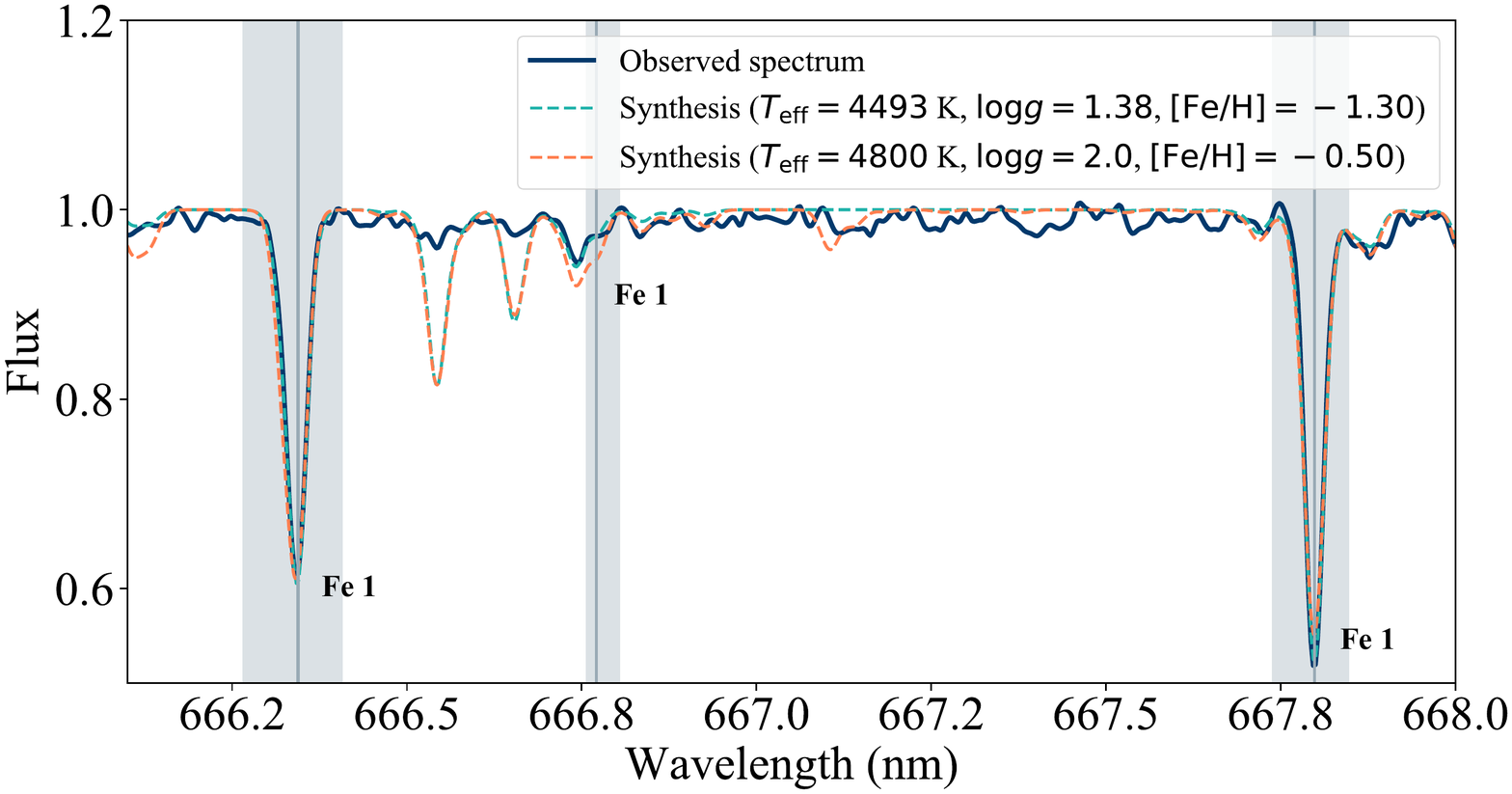}
\includegraphics[width=9cm]{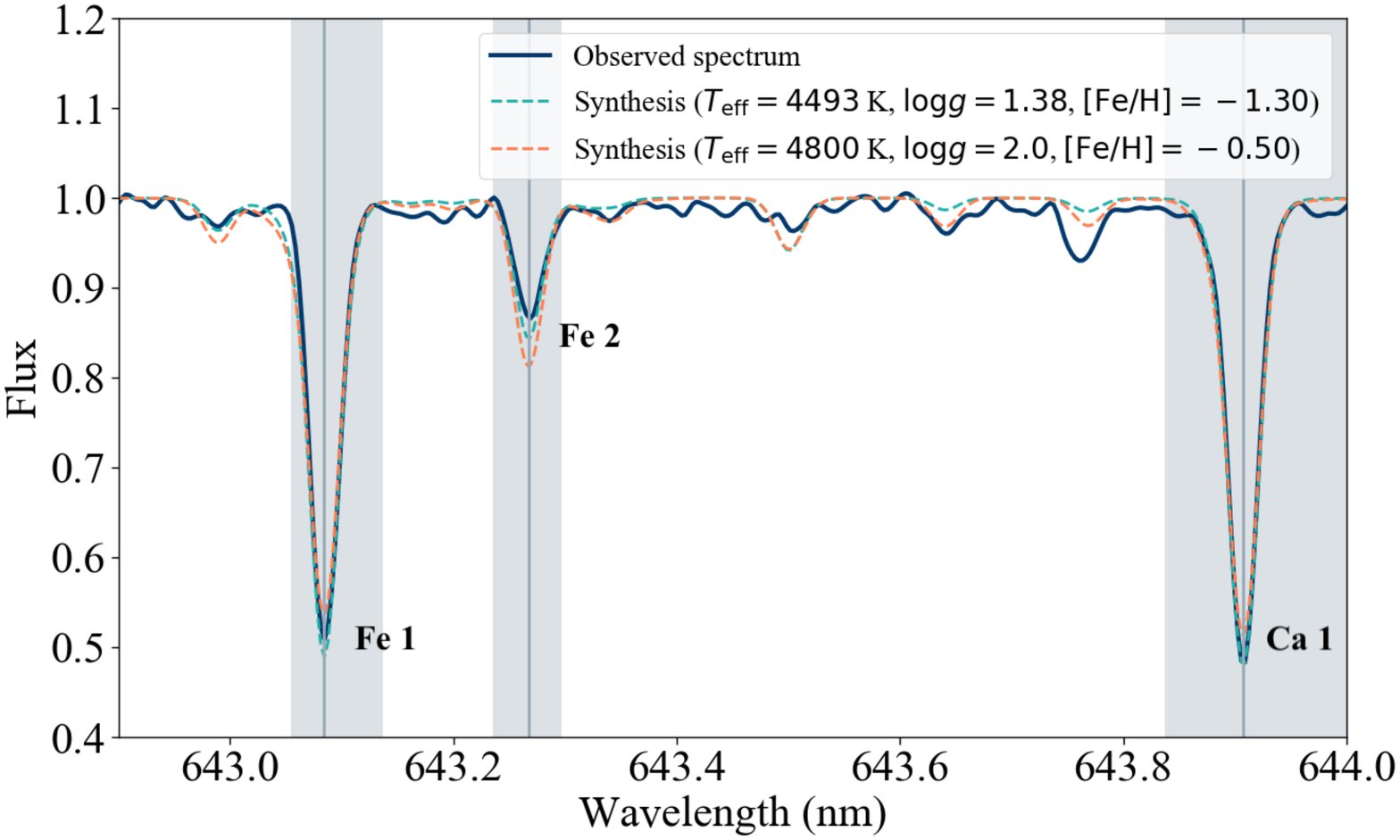}
\caption{Comparison between observed spectra (solid blue lines) and synthetic spectra (dashed green lines) based on our best fit stellar parameters. Dashed orange lines show the synthetic spectra built using stellar parameters from the literature. Top panel: region containing the magnesium triplet. Middle panel: region containing Fe 1 lines. Bottom panel: region containing Fe 1, Fe 2 and Ca 1 lines. The gray regions indicate the area of the different lines used as an example. The synthesis based on our best fit stellar parameters considers a star of $T_\mathrm{eff} = 4493$ K, $\log g = 1.38 $ and $\mathrm{[Fe/H]} = -1.30$. The synthesis based on literature stellar parameters considers $T_\mathrm{eff} = 4800$ K, $\log g = 2.0 $ and $\mathrm{[Fe/H]} = -0.5$.}
\label{fig:example_fit}
\end{figure}

In order to determine atmospheric parameters we adopted a similar procedure as the one implemented in \cite{casamiquela2019}.
We considered the line list from the Gaia-ESO survey \citep{heiter2015atomic}, which includes atoms and molecules, 
and the Grevesse solar abundances \citep{grevesse2007}. Within {\tt iSpec} we chose
the 1D atmospheric models \textsc{MARCS7} \citep{gustafsson2008} and the LTE radiative transfer code \textsc{TURBOSPECTRUM} \citep{alvarez1997near,plez2012turbospectrum}. We considered in our analysis only the region between 480 nm and 660 nm, since bluer regions have lower signal to noise and also because \texttt{iSpec} has been largely tested in the adopted region. When {\tt iSpec} uses \textsc{TURBOSPECTRUM}, it considers the technique of fitting synthetic spectra to the observations, in specific regions defined by the user, in this case those listed in Table~\ref{tab:regions_stellarparam} in the Appendix~\ref{line_regions}.  The atmospheric parameters are with this method determined by fitting on-the-fly the selection of spectral regions simultaneously until a good match between the synthesis and the observation is achieved (following a $\chi^2$ minimisation procedure). After the atmospheric parameters are decided, the chemical abundances can be determined line-by-line, using syntheses with the same radiative transfer code. The lines used to derive chemical abundances are also found in the Appendix~\ref{line_regions}, in Table~\ref{tab:regions_chemical}.  After generating the synthetic spectra, we visually verified the good agreement between it and the observed spectra. In Figure \ref{fig:example_fit}, we show a comparison between the synthetic spectrum built with our best fit stellar parameters (green dashed lines) and the observed spectrum (solid blue lines). We also show the synthetic spectrum built using the stellar parameters from \cite{massey2018runaway} (dashed orange lines). There is a good agreement between the observed spectrum and the synthetic one based on our best fit stellar parameters in the gray regions, which are those used for performing the fitting. The regions chosen to display the agreement between synthesis and observations focus on the Mg triplet at 517 nm,  and other regions in which both neutral and ionised iron lines of different strengths and excitation potentials fall. We note that the synthetic spectrum built using the stellar parameters presented in this work agree better with the observed one than the synthetic spectrum built using stellar parameters from the literature. Only the regions of Fe 1 and Fe 2 were considered for the determination of parameters, relying on the method of ionisation and excitation balance \citep{gray2005observation}: effective temperature does not need to depend on the strength or excitation potential of Fe 1 lines, and the metallicity needs to give the same result for Fe 1 and Fe 2 lines. This is achieved by adjusting the value of surface gravity, which depends on the strength of Fe 2 lines. The Mg triplet, which in this case was not used for the fitting, is displayed here as diagnostics, since its wings are very dependent on surface gravity. The good agreement between the synthetic spectrum based on our parameters and the observed spectrum in these regions indicates that $T_\mathrm{eff}$, $\log g$ and [Fe/H] are consistently calculated, addressing possible degeneracies in the parameters. In our analysis we did not include H lines because they suffer from strong non-LTE and 3D effects \citep{amarsi2018effective}, but in Appendix \ref{sec:Hlines} we give further discussions on the H line profiles for completeness.

With the aim of validating our atmospheric parameters, we considered a control sample of metal-poor stars studied by Car21 and for which we have parameters from the \textsc{APOGEE} survey. In this case, we used \textsc{APOGEE} $T_\mathrm{eff}$, $\log g$ and [Fe/H] as reference values, testing the accuracy of our data as well as estimating external uncertainties. We chose not to compare our stellar parameters with the ones presented in Car21, because the analysis done in that work is not fully spectroscopic.

\begin{figure}
\centering
\includegraphics[width=9cm]{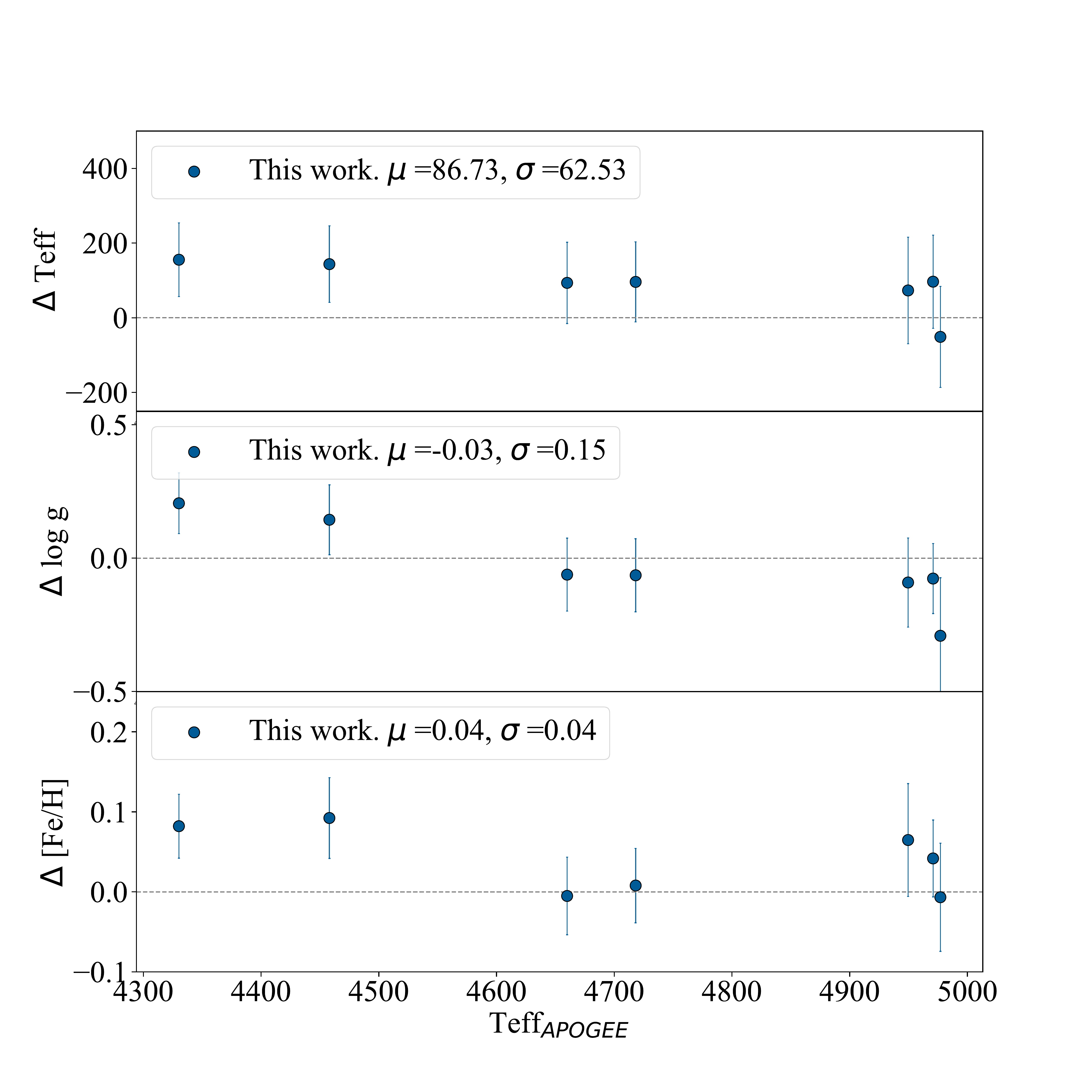}
\caption{Comparisons of our atmospheric parameters with APOGEE. Top panel: difference in effective temperature. Central panel: Difference in surface gravity. Bottom panel: difference in metallicity. The mean $\mu$ as well as the standard deviation $\sigma$ of each distribution is included in the panels. The uncertainties are computed as the quadratic sum of uncertainties from the pipeline and from the literature value.}
\label{fig:validation_dreia}
\end{figure}

In Figure \ref{fig:validation_dreia} we compare our results for the control sample. We plot the difference between the atmospheric parameters T$_\mathrm{eff}$, $\log g$ and [Fe/H]
obtained in our work and values from \textsc{APOGEE} survey. The mean of the differences between the results and the standard deviation are indicated in each panel for reference.

For all stars, there is a consistency between T$_\mathrm{eff}$ obtained by us and T$_\mathrm{eff}$ from \textsc{APOGEE} within 1 sigma. Nevertheless, we observe that our T$_\mathrm{eff}$ tend to be higher than the \textsc{APOGEE} results by $87\pm63$ K. In regard to $\log g$, all values obtained in this work agree with \textsc{APOGEE} values within 2 sigma. The mean difference of log g is -0.03 and the standard deviation is 0.15. Regarding the metallicity, there is an agreement within 2.2 sigma when comparing measured values from our work and \textsc{APOGEE} values. 
The mean difference for the metallicity is 0.04 dex, the standard deviation is 0.04 dex, thus there is a slight offset in which  we obtain metal-richer results than \textsc{APOGEE}. 

We obtained $T_\mathrm{eff} = 4493 \pm 102$ K, $\log g = 1.38 \pm 0.15$ and $\mathrm{[Fe/H]} = -1.30 \pm 0.1$ for J01020100-7122208. Here we considered as the uncertainty of our measurements the quadratic sum of the internal value provided from our pipeline and the standard deviation from the comparison presented in Figure \ref{fig:validation_dreia}. Our parameters point towards the star being a metal-poor red giant star, even when considering the systematic offset obtained with respect to APOGEE, which agrees with Mas18 but not with Neu18. In any case, our study indicates that the star is more metal-poor then previously reported, and is consistent with the star being a member of the stellar halo.   

\subsection{Kinematics}

We calculated the velocities $(U,V,W)$ for J01020100-7122208 using the \textsc{astropy} package \citep{robitaille2013astropy,price2018astropy}. We adopted a Galactic height of $z_\odot = 0.0025$ Kpc  \citep{juric2008milky} and Galactic radius $R_\odot = 8.2$ Kpc \citep{mcmillan2016mass} for the Solar position, and $(U, V, W)_{\odot} = (11.10, 247.97, 7.25)$ km s$^{-1}$ for the Solar velocity relative to the Galactic Center, following \cite{matsuno2020star}.
We used right ascension, declination and proper motions from Gaia EDR3 and the RVs obtained from our \textsc{MIKE} spectra. We adopted $U$ positive toward Galactic Center, $V$ positive in the direction of Galactic rotation and $W$ positive toward the North Galactic Pole. 
The final velocity in the Galactic rest frame is $(U,V,W) = (-185.26,-171.64,-159.30)$ km s$^{-1}$.  This indicates that the star is counter-rotating, e.g. it is part of the retrograde halo, which is consistent with the metal-poor nature of the star.

\subsection{Distance and other stellar properties}

We use the \textsc{minimint} package \citep{sergey_koposov_2020_4002972} that relies on the MIST library of synthetic isochrones \citep{dotter2016mesa,choi2016mesa} to map the mass, age and metallicity to absolute magnitudes in a variety of filters as well as surface gravity and effective temperature. We then model the observed photometry, Gaia parallaxes, and spectroscopic stellar atmospheric parameters similarly to \citet{koposov2020}. The model parameters are the mass, age, metallicity, extinction, distance to the star and an additional systematic photometric scatter that is added in quadrature to all the magnitude uncertainties. We use the optical photometry from \textit{Gaia} EDR3 \citep{riello2021} and Skymapper DR2 \citep{onken2019} and IR photometry from 2MASS \citep{skrutskie2006} and WISE surveys \citep{eisenhardt2020}. As the parallax solution of \textit{Gaia} is known to have systematic spatially dependent biases \citep{lindegren2021} we also introduce an additional parallax offset parameter with a $\delta \omega/(1 {\rm mas}) \sim N(0,0.02)$ prior.  We use the Nested Sampling algorithm \textsc{MULTINEST} \citep{feroz2008multimodal} as implemented in Python by \citet{buchner2014x} to sample the posterior.  Similarly to \citet{koposov2020} we run the model in two configurations, one where the photometry alone is modeled, and a second where we use the stellar atmospheric parameters in the fit together with photometry.

\begin{table}
\caption{Result of Spectrophotometric Analysis}
\label{tab:spectrophoto}
\centering 
\begin{tabular}{c c c}
\hline\hline
Property & Photometric & Spectrophotometric \\ \hline
Mass ($M_\odot$) & $3.39\pm0.25$ & $1.09\pm0.10$ \\ 
Age ($\mathrm{Gyr}$) & $0.244\pm0.041$ & $4.507\pm1.437$ \\      
Distance ($\mathrm{kpc}$) & $10.39\pm1.17$ & $8.74\pm0.74$ \\
Extinction ($\mathrm{mag}$)& $0.102\pm0.017$ & $0.083\pm0.016$ \\
Systematic error ($\mathrm{mag}$) & $0.028\pm0.007$ & $0.053\pm0.015$\\
Parallax offset ($\mathrm{mas}$) & $0.0034\pm0.0122$ & $-0.0092\pm0.0146$ \\
\hline 
\end{tabular}
\end{table}

\begin{figure*}
\centering
\includegraphics[width=17cm]{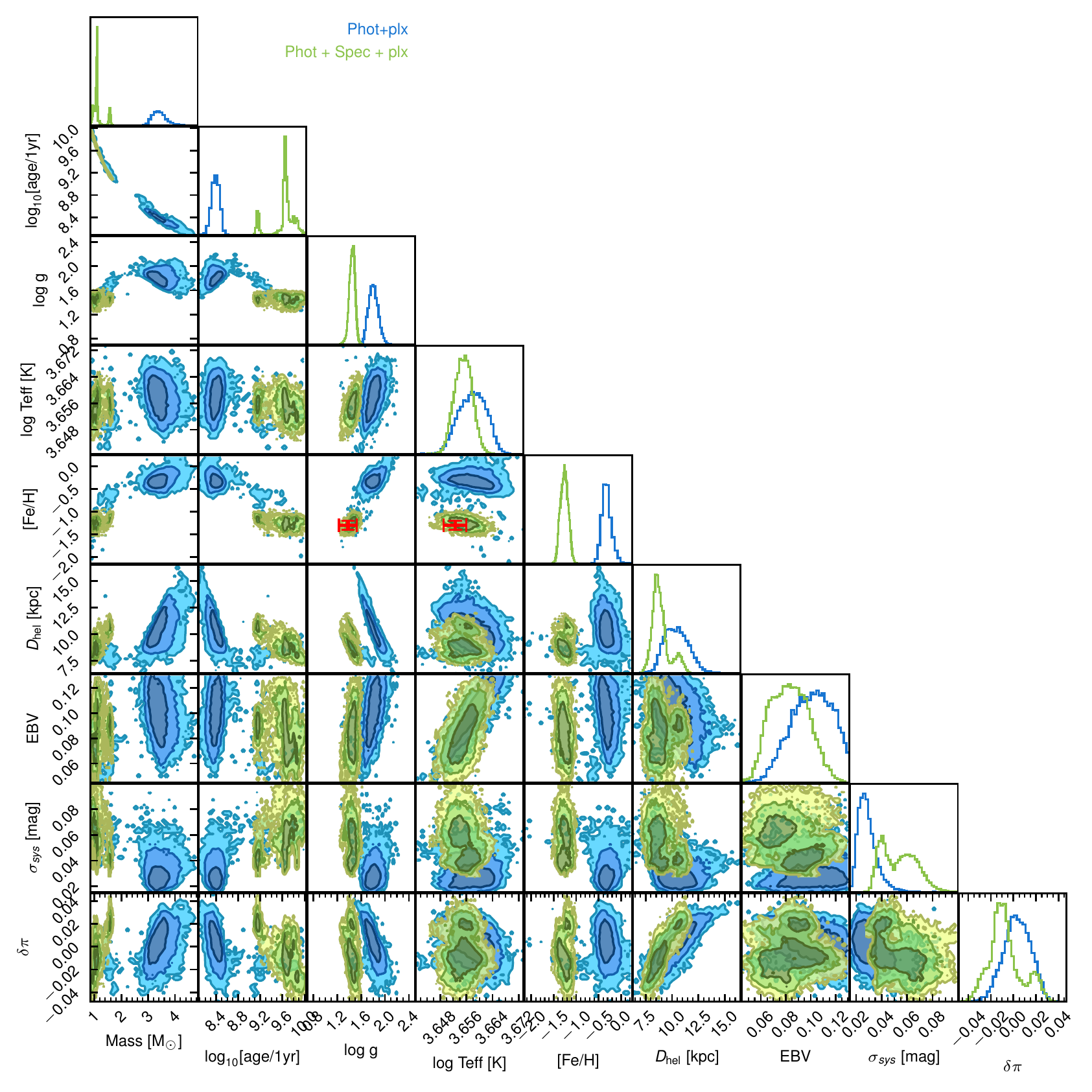} \\
\caption{Corner plot of the posterior from our photometric (blue) and spectrophotometric (green) analyses of the star. The red point is the measurement of the atmospheric properties derived from the spectrum.}
\label{fig:ages}
\end{figure*}

We show the posterior of our photometric and spectrophotometric analyses in Fig. \ref{fig:ages} and the corresponding means and standard deviations of the posterior samples in Table \ref{tab:spectrophoto}. These two analyses are in some tension, with the photometric analysis favouring an interpretation of the star as a young, metal-rich, high-mass giant star while the spectrophotometric analysis favours it to be an old, metal-poor, low-mass giant star. We note that the photometric-only analysis implies values of $T_\mathrm{eff}$, $\log g$ and $[\mathrm{Fe}/\mathrm{H}]$ which are entirely inconsistent with the values we measured from the spectra. 

We note that the results of photometric data only are closer to those of \cite{massey2018runaway}. Considering that in that case both analyses use \texttt{Gaia} data and were based on the same technique, this is expected. Our high resolution analysis however points towards a star that is significantly more metal-poor than previously believed. It is well-known that metalliticies can have a large impact on stellar models \cite[e.g.]{tayar2017}, causing a large systematic effect if solar-metallicities are considered for metal-poor stars. 
We see here how a high resolution spectrum can provide additional useful information. Furthermore,  in Appendix \ref{sec:Hlines} we verified from the hydrogen lines that our spectroscopic results  are consistent with those line profiles (what was also illustrated in Figure \ref{fig:example_fit}), hence confirming that a spectrophotometric analysis in this case is needed.

Therefore, our discussion will assume that this star is indeed an old, metal-poor, low-mass giant star, since it is the result obtained when we use more complete information about J01020100-7122208.

\subsection{Orbit calculation}
We characterised the full variety of possible orbits of this star through the Galaxy by drawing random values from our posterior on the present day distance, proper motion and radial velocity and integrating them forwards and backwards in time. The spectrophotometric analysis above yields a posterior on the distance to the star and we can sample the radial velocity from the normal distribution given by our measured mean and uncertainty. However, more work was needed to obtain samples of the proper motions.

The spectrophotometric analysis of the previous section yields a posterior on the distance to the star which is both more precise and shifted relative to the parallax reported by \textit{Gaia}. The uncertainty in \textit{Gaia}'s measurement of a star's parallax is correlated with the uncertainty in the proper motion, meaning that any extra information on a star's true parallax will also cause our estimate of the star's proper motion to change. The implication of this is that randomly sampling the proper motions from the \textit{Gaia} reported means and uncertainties will yield proper motions which are inconsistent with our best estimate of the distance. 

\textit{Gaia} describes the correlation between the parallax and proper motions through a trivariate normal distribution with mean $\boldsymbol{m}=(m_\varpi,m_{\mu\alpha\ast},m_{\mu\delta})^\intercal$ and covariance matrix $\mathbf{S}$. The spectrophotometric posterior on the distance cannot in general be described as a normal distribution, so we instead treated each posterior sample as giving a point estimate of the true parallax $\tilde{\varpi}$. We then conditioned the proper motion distribution on each of those values in turn to derive updated estimates of the mean $\tilde{\bf{m}}_\mu$ and covariance matrix $\tilde{\mathbf{S}}_{\mu\mu}$ of the bivariate normal distribution on the proper motions, through the equations
\begin{equation}
    \tilde{\boldsymbol{m}}_{\mu} = \boldsymbol{m}_{\mu} + \frac{\tilde{\varpi}-m_\varpi}{\sigma_\varpi^2}\mathbf{S}_{\mu\varpi},
\end{equation}
\begin{equation}
    \tilde{\mathbf{S}}_{\mu\mu} = \mathbf{S}_{\mu\mu}-\frac{1}{\sigma_\varpi^2}\mathbf{S}_{\mu\varpi}\mathbf{S}_{\varpi\mu},
\end{equation}
where quantities like $\mathbf{S}_{ab}$ refer to the corresponding submatrices. We drew one sample from each of these distributions, such that for each distance sample from the spectophotometric posterior we had a consistent proper motion sample.

For 2,500 of these samples, we integrated the orbit of the star forward and backward in time for $1\;\mathrm{Gyr}$ using the \textsc{gala} \citep{gala} Python package, assuming the default Milky Way potential and transforming into the Galactocentric coordinate frame using the latest \textsc{astropy} recommended values \citep{reid2004,gravity2018,drimmel2018,bennett2019}.

\begin{figure*}
\centering
\includegraphics[width=15cm]{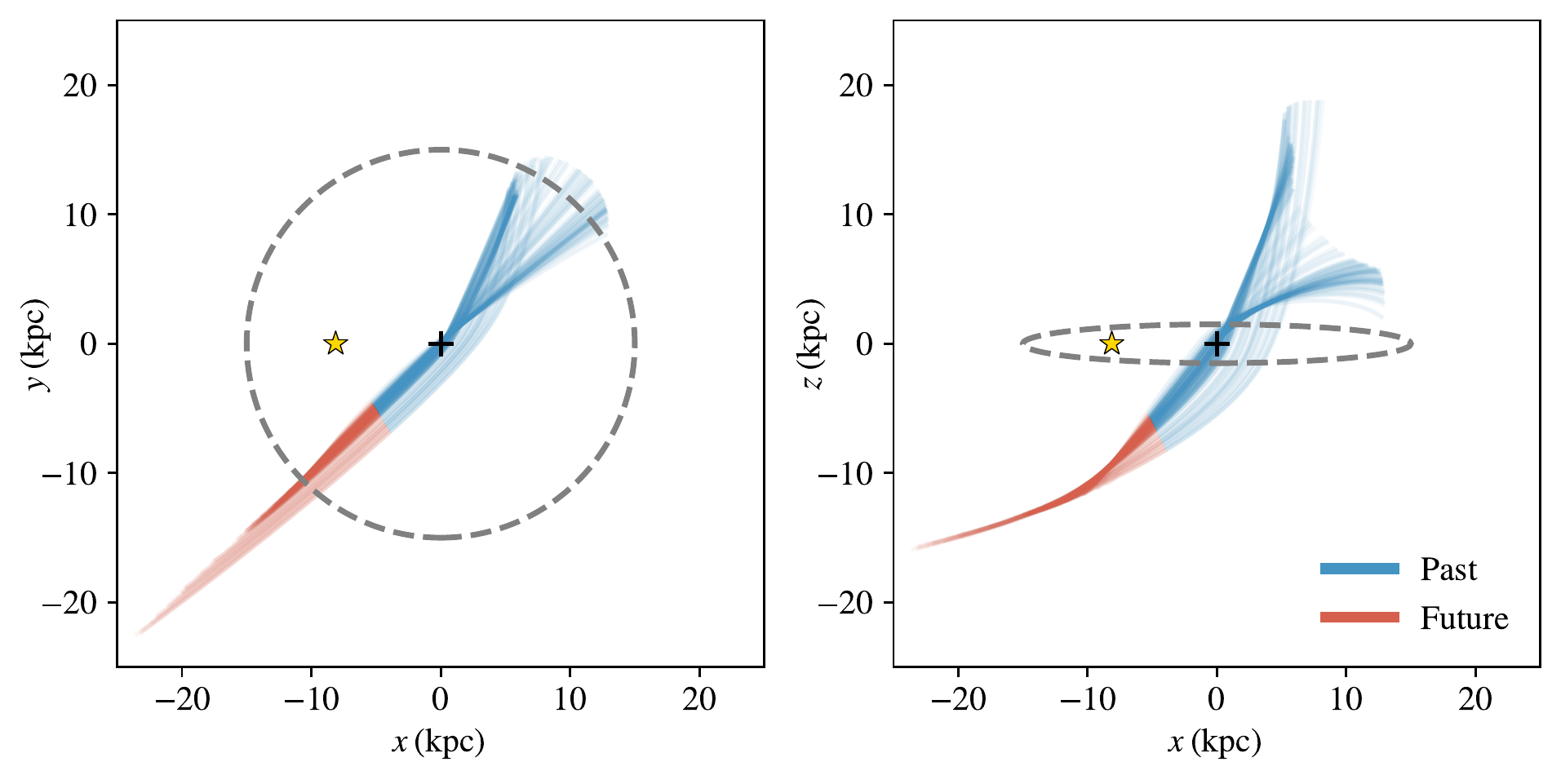} \\
\caption{A random sample of 250 possible past and future trajectories of the star through the Galactic potential. The location of the Sun is marked with a yellow star, the Galactic centre with a cross and the Milky Way disk with a dashed ellipse.}
\label{fig:orbits}
\end{figure*}

We show in Fig. \ref{fig:orbits} a random subsample of 250 orbits out of the 10,000 orbits that we integrated. While the uncertainty in the present day distance to the star (in particular the second mode visible in Fig. \ref{fig:ages}) produces a spread of trajectories, the main bulk of the trajectories pass close to the centre of our Galaxy. To test the hypothesis that this star originates in the Galactic centre, we identified where each past trajectory last crossed the Galactic disk, finding that the star last crossed the plane at a distance of $550_{-361}^{+1129}\;\mathrm{pc}$ from the Galactic centre and that the crossing location was at least $113\;\mathrm{pc}$ from the Galactic centre with 99 per cent confidence. Another way of expressing this is that the star's eccentricity is $0.917_{-0.025}^{+0.005}$, making this star's orbit highly eccentric but not perfectly radial. If we repeat the orbit integration with the photometric analysis, then we find that the star last crossed the plane at a distance of $2673_{-1721}^{+2329}\;\mathrm{pc}$ from the Galactic centre and that the crossing location was at least $135\;\mathrm{pc}$ from the Galactic centre with 99 per cent confidence. Neither of the analyses support the hypothesis that the star originated in the Galactic centre.

\section{Chemical abundances}\label{sec:results}

\begin{figure*}
\centering
\includegraphics[width=18cm]{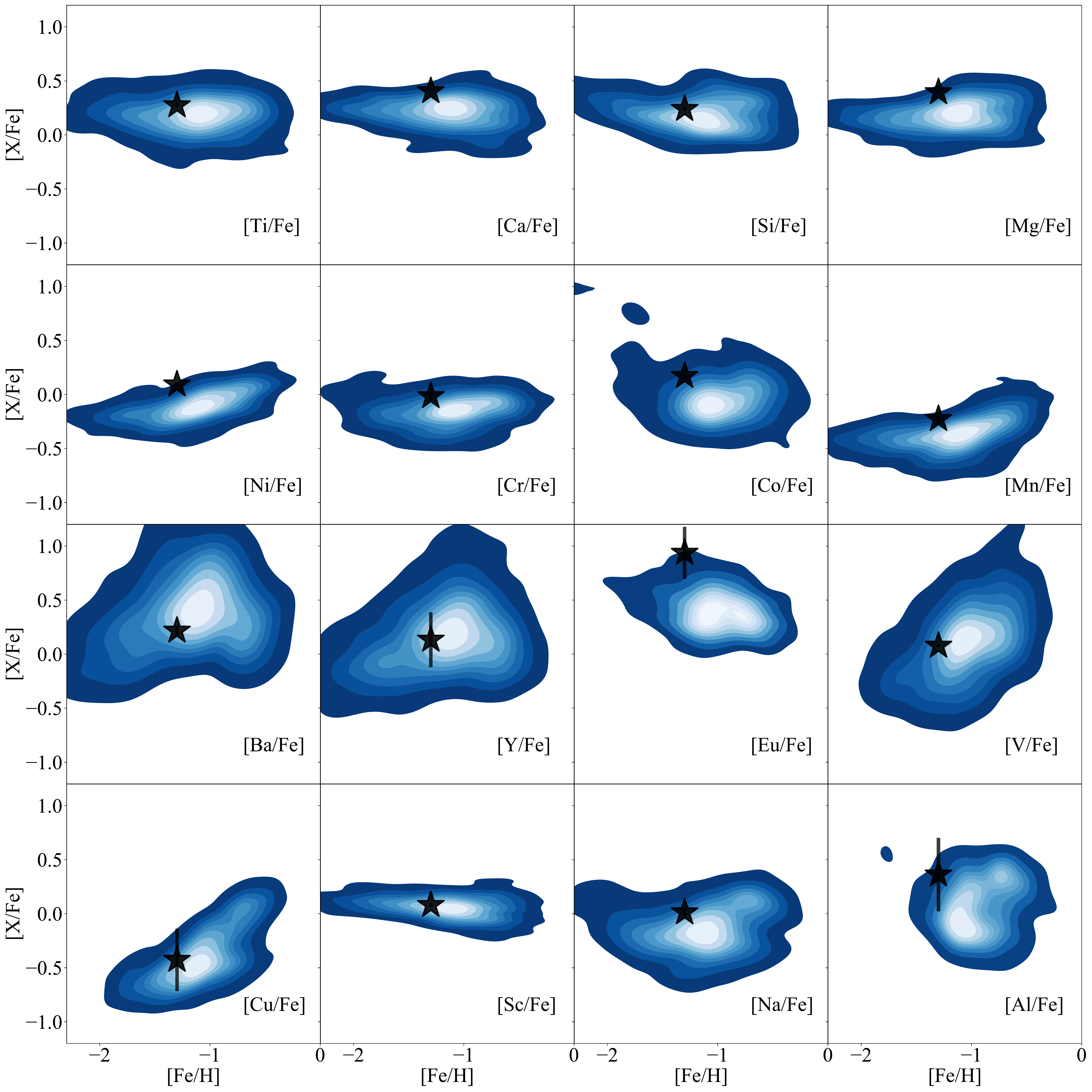} \\
\caption{Chemical abundances. Black star refers to J01020100-7122208.
Blue contours refer to a kernel density estimation of retrograde stars from  \textsc{GALAH} survey used here as a reference sample.}
\label{abundance_space}
\end{figure*}

The chemical abundances  can be seen in Figure \ref{abundance_space}, where we show J01020100-7122208 as a black filled star. In each panel we display the measured abundance ratio [X/Fe] of a different element. In the same figure, we adopted as reference a sample of retrograde stars from the \textsc{GALAH} survey \citep{buder2020galah+}, represented as contours of a kernel density estimation. The control sample was taken from the DR3 of \textsc{GALAH}, selecting only stars with T$_\mathrm{eff}$ $\leq$ 5500 K and $\log g$ $\leq$ 2.0. We also did a quality selection, where we chose only stars with flag\_sp = 0 (i.e. no problems with the determination of parameters), flag\_fe\_h = 0 (i.e. no problems in the determination of [Fe/H]) and flag\_X\_fe = 0 (no problems in the determination of the abundance [X/Fe]). Further details can be found in \cite{buder2020galah+}.

For all these stars we used Gaia positions, proper motions and RVs to calculate the respective total velocities.  We selected only stars with  $V \leq 0$ km s$^{-1}$. Since J01020100-7122208 is a retrograde star, we compared its chemical composition with other stars to see if they come from the same Galactic component. In general we find that the chemical composition of J01020100-7122208 is very similar to the GALAH retrograde stars. 
The results of individual chemical elements is discussed in more detail below.

\subsection{\texorpdfstring{$\alpha$}{} elements}

$\alpha$ elements are those formed by the capture of $\alpha$ particles in the core of stars during post-main sequence burning and are dispersed in the interstellar medium (ISM) mainly by core collapse Type II supernovae (SNII) (e.g. \citealt{timmes1995galactic},  \citealt{kobayashi2006galactic},  \citealt{nomoto2013nucleosynthesis}). In this work we explore the following $\alpha$ elements: calcium (Ca), silicon (Si) and magnesium (Mg). We also explore titanium (Ti), that according to nucleosynthetic models is not produced by the $\alpha$-capture channel but the [Ti/Fe] distribution as a function of [Fe/H] behaves like other $\alpha$ elements. 

Following the time-delay model \citep{tinsley1979stellar,matteucci1986relative}, there is a delay in the ejection of Fe-peak elements such as Fe and Mn by Type Ia supernovae (SNIa) when we compare it with the fast ejection of $\alpha$ elements by SNII. The result of the time-delay enrichment of Fe, for example, is that the oldest stars are both $\alpha$-rich and metal-poor. This class of elements are also an indicator of the star formation rate of a progenitor galaxy: the richer a star is in $\alpha$-elements, the higher the star formation rate of the progenitor galaxy \citep{gilmorewyse98}.

The top row of Figure \ref{abundance_space} shows the abundance ratios of $\alpha-$elements and it is possible to see that J01020100-7122208 is an $\alpha$-rich star ([$\alpha$/Fe] $\approx$ 0.30). This enhancement is consistent with the star being old. It is further seen that J01020100-7122208's [$\alpha/$Fe] has abundance ratios consistent with the retrograde GALAH stars. 

\subsection{Iron-peak elements}

Iron-peak elements are formed by several different nucleosynthetic channels, but are mainly dispersed into the ISM by SNIa \citep{iwamoto99}, in the same way as iron.  In this work we explored four iron-peak elements: nickel (Ni), cobalt (Co), chromium (Cr) and manganese (Mn) seen in second row of Figure \ref{abundance_space}. Cr is a chemical element that follows the behaviour of iron, which is seen in Figure \ref{abundance_space}, where the abundance of this element is similar for all stars. On the other hand, Mn varies with the metallicity and is a good trait to distinguish populations. Mn for example is a very good tracer of SNIa because it is more produced in SNIa than SNII in relation to Fe \citep{kobayashi2009role}.

In Figure \ref{abundance_space}, we see that the star is in the Mn-rich part of the diagram, indicating that among the retrograde halo population stars, it is likely not part of the oldest stars of this component. In the same figure, we can see that Co does not vary with the metallicity, while Ni shows a slight increase with the metallicity. Also, from the control sample we can see that there is a Ni-rich population of stars in this Galactic component, but J01020100-7122208 is not part of this sub-population.

\subsection{Neutron-capture elements}

Neutron capture elements can be divided in two subclasses: the r-process elements and the s-process elements. In this work we studied the s-process elements yttrium (Y) and barium (Ba) as well as  the r-process element europium (Eu). S-process elements are produced by low- to intermediate-mass AGB stars \citep{busso1999nucleosynthesis}, while Eu is believed to be produced mostly in neutron star mergers \citep{matteucci2014europium}. The abundances of Y and Ba measured in J01020100-7122208 agree well with our control sample. J01020100-7122208 shows an overabundance of europium (Eu), having a measured value of 0.93 $\pm$ 0.24. Both \cite{matsuno21} and \cite{aguado20} reported that accreted stars from a progenitor galaxy named Gaia-Enceladus \citep{Belokurov2018,Helmi2018} shows an overabundance of this element, with a central distribution of [Eu/Fe] $\approx$ 0.5, but also containing stars with higher abundances. The overabundance of [Eu/Fe] is evidence that J01020100-7122208 may be an accreted star from Gaia-Enceladus. 

\begin{figure*}
\centering
\includegraphics[width=17cm]{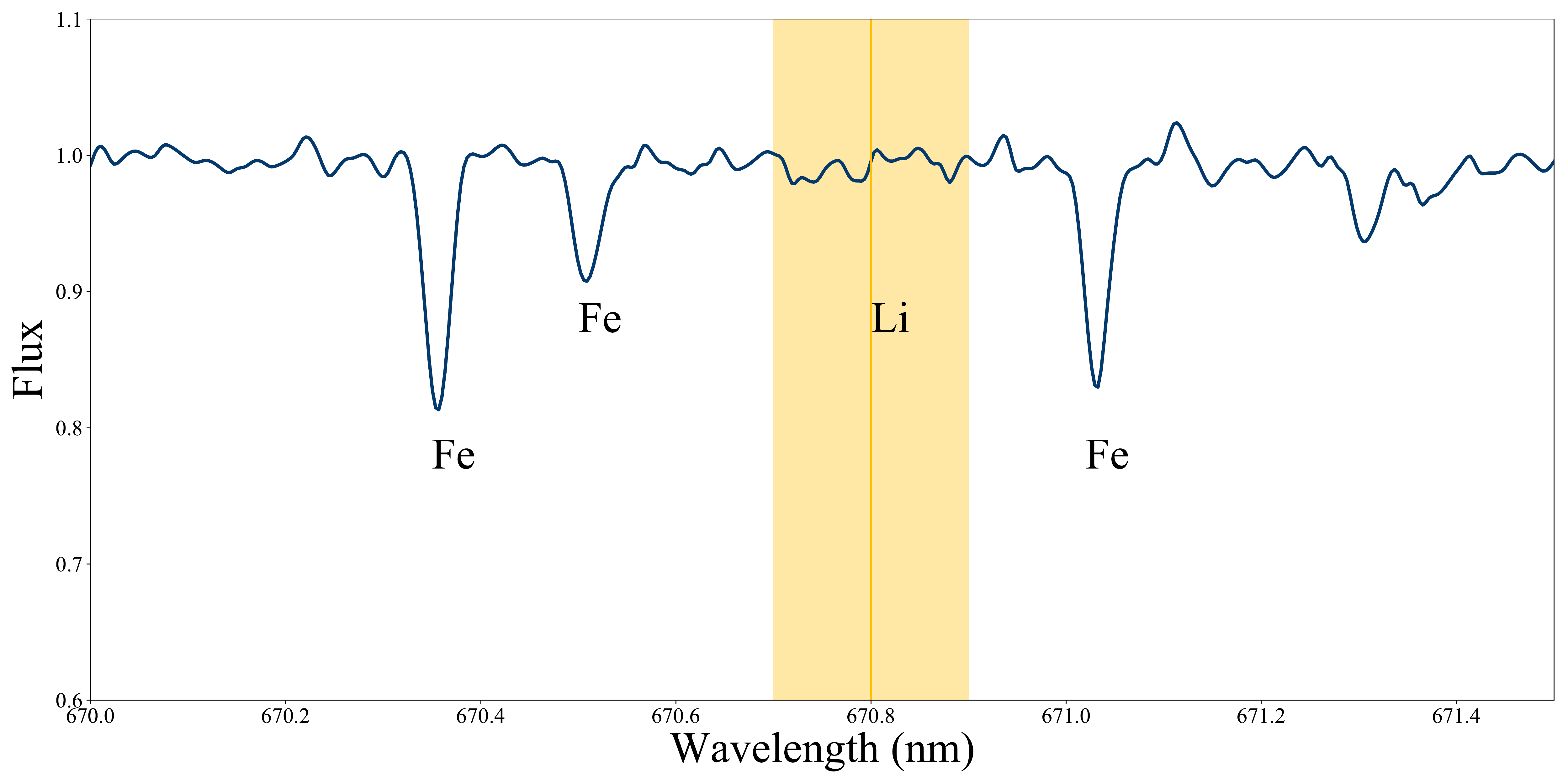}
\caption{Spectrum of J01020100-7122208 near the region of the lithium line at 670.7 nm. The vertical yellow line and region represents where the Li line should be if observed.}
\label{fig:lithium}
\end{figure*}

\subsection{Odd-Z elements}

In this work we measured chemical abundances of five odd-Z elements: sodium (Na), aluminum (Al), copper (Cu), vanadium (V) and scandium (Sc). Na and Al are thought to be produced by SNII, having a production dependent on the abundance of C and N in the environment \citep{kobayashi2006galactic} as well as with the metallicity of the progenitor  (e.g. \citealt{das2020ages}).
 
They are also believed to be produced by AGB stars \citep{nomoto2013nucleosynthesis}. From Figure \ref{abundance_space}, we can see that the abundances of these two chemical elements agree well with the chemical abundances of retrograde halo stars.  While Cu is thought to be formed through different channels including week s-process, SNIa, hypernovae and by massive stars \citep{pignatari2010weak,nomoto2013nucleosynthesis}, V is through to be produced mainly in explosive silicon burning in SNII and Sc is believed to be produced in carbon and neon burning phases in massive stars \citep{woosley1995evolution}.
In Figure \ref{abundance_space} we can see that the
abundances of these three elements are compatible with the typical abundance of retrograde halo stars.

\subsection{The absence of lithium}

Lithium (Li) is a chemical element that can correlate with the age of stars \citep[and references therein]{randich2021light}. Li is burned very fast in stars and the intensity of depletion is related with the mass of the star, with the main depletion process occurring even before the main sequence.
The depletion of Li in massive stars is not relevant, but it is very important in low-mass stars, independent of the metallicity. Therefore, regular metal-poor red giants stars do not show the presence of Li.  From our spectrum, we could not detect Li, which is further evidence of the hypothesis that this star is a regular old red giant. We present in Figure \ref{fig:lithium} the spectrum of J01020100-7122208 in the region near the lithium line at 670.7 nm. The position of the line is indicated in yellow and shows no line.

\section{The origin and nature of J01020100-7122208 }\label{sec:discussion}

In contrast to what was reported by Neu18 and Mas18, here we find that J01020100-7122208 is a low-mass (1.09 $\pm$ 0.10 M$\odot$) intermediate-age (4.51 $\pm$ 1.44 Gyr) metal-poor ([Fe/H] = -1.30 $\pm$ 0.10) red giant star. Both Neu18 and Mas18 concluded this star was young. Neu18 determined an age of 30 Myr, while Mas18 obtained an age of 180 Myr. This is a consequence of the rather high masses determined in these works, of 9 and 3-4 M$\odot$, respectively.

The main conclusion of Mas18 was that this star was likely an object accelerated by the central black hole of the Milky way. Here, with improved astrometric data from Gaia EDR3 combined with a detailed high resolution spectral analysis, we found instead that the star probably passed about 550 pc away from the central black hole, which is too far for it to have been  ejected from the centre.

Furthermore,  Mas18 reported a metallicity of [Fe/H] = -0.5, while we obtained a value of [Fe/H] = -1.30 $\pm$ 0.10. In Mas18, their metallicity and age led them to conclude that this star originated in the Milky Way disk. In our work, we found that the star has a retrograde velocity, which, together with its low metallicity and high [$\alpha$/Fe] ratio, make it likely to belong to the halo. Here, however, our age is not the typical age of halo stars \citep[of approximately 10~Gyr, ][]{jofre2011age, kalirai2012age,das2020ages}. Further  discussions are found below. 

\subsection{Chemistry}

From the atmosphere of low-mass stars it is possible to measure chemical abundances that imprint the chemical composition of the molecular cloud from which the star was born. This is assuming that stellar internal processes do not significantly impact the chemical elements in the surface of the star. Therefore, by examining the chemical composition of this star, we can shed light about its origin. We considered $\alpha-$ elements (Ca, Si, Mg and Ti), iron-peak elements (Ni, Cr, Co and Mn), odd-Z elements (Na, Al, Cu, V, Sc) and neutron-capture elements (Y, Ba and Eu).

From Figure \ref{abundance_space} it was possible to see that the chemical abundances of J01020100-7122208 agree well with the distribution of retrograde stars, which corroborates the hypothesis that this star, although fast, is a member of this component. It is believed that the Milky Way retrograde halo is partially formed by stars accreted from other galaxies \citep{gratton2003abundances, carollo2007two}.
Therefore, J01020100-7122208 might be one such star that was formed in another galaxy and was accreted later into the Galaxy.

As mentioned before, $\alpha-$elements are good tracers of star formation rate. A star with enhanced values of [$\alpha$/Fe] is evidence that it was born in an environment were the star formation rate was high. In the case of low-mass dwarf galaxies, we have an environment where the star formation rate is lower than in the Milky Way \citep{tolstoy2009star,nissen2010two}.  Therefore, it is expected that stars born there are more $\alpha$-poor than stars born in situ \citep{nissen2010two}. But if the progenitor galaxy was relatively massive (of the order of $10^{9-10}$ M$_\odot$), the values of $\alpha$ elements in the low metallicity regime can be similar to the one observed in metal-poor stars of the Milky Way \citep{das2020ages}.  J01020100-7122208 is a star of high [$\alpha$/Fe] abundance ratios  compared to the rest of the counter-rotating halo stars from GALAH (see Fig.~\ref{abundance_space}). Thus, if accreted, the progenitor galaxy should have been massive. In particular, it is possible that the progenitor was Gaia-Enceladus \citep{Belokurov2018, Helmi2018}, with a mass of $\approx$ 10$^{9}$ M$_\odot$ \citep{vincenzo2019fall,das2020ages,feuillet2020skymapper}. 
We also note that chemical abundances of J01020100-7122208 agree well with the chemical pattern of Gaia-Enceladus stars analysed by \citealt[][and Car21]{matsuno2019origin}.

\subsection{Age}
The high $\alpha$ abundance together with the low metallicity are strong evidence that this star is old \citep{schuster2012two,hawkins2014relative,montalban2021chronologically}. The fact that the star has a low abundance of Mn reinforces this suggestion. This star however is not as old as other regular halo stars. 

\cite{chiappini2015young} reported few CoRoT \citep{baglin2006scientific} red-giant stars that despite being metal-poor and enhanced in [$\alpha$/Fe], are apparently young (younger than 7 Gyr approximately). In that work, the authors proposed that they  formed near the bar co-rotation, which is a region where gas can be kept inert for longer times than in other places of the Galaxy \citep{bissantz2003gas,combes2013gas}. These {\it young $\alpha$-rich} (Y$\alpha$R) stars were also reported in \cite{martig2015young}, where the authors found 14 of these objects in Kepler \citep{borucki2010kepler}, where 5 of them had ages below 4 Gyr. 

Currently, there are two main explanations for the Y$\alpha$R stars, though neither are fully conclusive.  The first scenario, mentioned above \citep{chiappini2015young,zhang2021high} is that the stars are truly young and have originated near the Galactic center and migrated outwards. The other scenario  \citep{jofre2016cannibals, yong2016graces} postulates that these stars were binary stars which experienced mass transfer, e.g. they are evolved blue stragglers.  Hence, the ages were probably not calculated correctly and these stars are not really younger than typical old stars. The challenge is that not all such stars show evidence of binary evolution \citep{aguirre2018confirming,hekker2019origin}.
Given that the elemental abundances and kinematics of such stars mimics those of old stars \citep{matsuno2018optical}, it is very difficult to explain their origins being different than normal thick disk or halo stars.  On the other hand, considering the poor evidence of variation in RV and no obvious high rotation from the line profiles \citep{jofre2016cannibals,aguirre2018confirming} it is not necessary that J01020100-7122208 is in a binary system now. However, the binary could have merged \citep{izzard2018binary}. 

It is also possible that the star was part of a binary system in the past and when the companion exploded, J01020100-7122208 was ejected with higher velocity. That scenario was proposed by Neu18 and Mas18. In both cases, the star might have been `rejuvenated' like a blue straggler, explaining our age determination of 4 Gyr.

As discussed in \cite{martig2015young} neither the uncertainties of the measured ages or the uncertainty related to their methodology are big enough to explain the young ages. Similarly, considering the uncertainty of the age measured in our work, we can not claim this star is as old as typical retrograde halo stars. In general, age determination is a challenging task. In the case of seismic ages (such as the ones reported in \citealt{chiappini2015young} and \citealt{martig2015young}), to calculate accurate ages of metal-poor giants is challenging because corrections are required in the scaling relations when studying metal-poor stars \citep{epstein2014testing} and also because since mass loss is not a very well known process, in the case of stars in the red clump, the mass loss experience is previous evolutionary phases can not be well accounted for (see \citealt{anders2017galactic,montalban2021chronologically}). In the same way, calculating ages of metal-poor giant stars using isochrone-fitting techniques (as done in this work) is also challenging since the isochrones tend to clump together in the region of the red giant branch (as commented in \citealt{martig2015young}).

\subsection{Kinematics}
We found that the star last crossed the Galactic plane at least $113\;\mathrm{pc}$ from the Galactic centre with 99\% confidence, arguing against acceleration by Sgr A$^\ast$. We note that we have not included the uncertainties in the Galactic potential or the Solar position and motion when calculating that confidence interval, and that it may be more plausible for this star's orbit to have passed through the Galactic centre if those are included. However, the null hypothesis must be that this star is an eccentric halo star, since the eccentricity, angular momentum and energy are fully consistent with that hypothesis.

A useful contrast can be found with the only star known to have been accelerated by Sgr A$^\ast$, S5-HVS1. \citet{koposov2020} followed a near-identical procedure to this work and found that the 90\% confidence region of the star's last crossing through the Galactic plane included the Galactic centre. However, the strongest evidence of the Galactic centre hypothesis was S5-HVS1's young age (less than $100\;\mathrm{Myr}$) and extreme velocity ($1755\;\mathrm{km}\;\mathrm{s}^{-1}$), which ruled out any other hypothesis. The star J1020100-7122208 is unlikely to have crossed the plane near the Galactic centre and is old and slow-moving, with a current velocity of $257\pm23\;\mathrm{km}\;\mathrm{s}^{-1}$. An extreme scenario is not required to explain its kinematics.

\subsection{Putting the puzzle together}

From the abundances alone, we conclude that if the star is an accreted object, the progenitor must have been relatively massive. In particular, it is possible that the progenitor is Gaia-Enceladus, since the chemical pattern of J01020100-7122208 is in good agreement with what was reported by \citealt[][and Car21]{matsuno2019origin} for stars from this progenitor. Another evidence that chemically supports the idea that the progenitor is Gaia-Enceladus is the overabundance of [Eu/Fe]. Based on the Toomre diagram presented in \cite{Koppelman2019} that shows the velocity distribution of stars from different progenitors including Gaia-Enceladus, and other dwarf galaxies like Sequoia, Thamnos 1 and Thamnos 2,  we see that J01020100-7122208 has a velocity profile consistent with Sequoia stars having $V = -171.64$ km s$^{-1}$ and $\sqrt{U^{2}+W^{2}} = 244.33$ km s$^{-1}$. 
However, J01020100-7122208 has a highly eccentric orbit (with an eccentricity of about 0.9). \cite{koppelman2020massive} reported that 75 per cent of Gaia-Enceladus stars have eccentricities higher than 0.8 (see also \citealt{mackereth2018origin}), 
while \cite{myeong2019evidence} reported that the typical eccentricity for Sequoia stars is 0.6. It is possible that the star comes from Gaia-Enceladus or Sequoia, although the chemistry is more consistent with Gaia-Enceladus. In any case, its origin is most likely extragalactic and therefore its younger age suggests it is rather an evolved blue straggler and not a younger star coming from the inner Galaxy. Y$\alpha$R stars there were likely accreted by the Milky Way have been reported before in the literature (e.g. \citealt{das2020ages}).

\section{Conclusion}\label{sec:conclusion}

J01020100-7122208 is a star which was reported as a serendipitous discovery a decade ago. Before the Gaia survey, this star was claimed to be a runaway yellow super giant from the Small Magellanic Cloud \citep{neugent2018runaway}. Immediately after Gaia Data Release 2, the star was reanalysed and claimed to be likely formed in the Milky Way disk and ejected to the halo by the central black hole of the Galaxy \citep{massey2018runaway}. This latter work warned that with newer Gaia data releases this conclusion could change. Therefore, now that Gaia EDR3 is available, the time is right to revisit this mysterious object.

In our work, we analysed J01020100-7122208 using astrometric, photometric and spectroscopic data to determine its age, orbit and chemical composition. It is the first time this star was chemically characterised. From our analysis we found that this star is a metal-poor red giant, with a metallicity of [Fe/H] = -1.30 $\pm$ 0.10. This metallicity value is lower than previously reported:  \cite{massey2018runaway} obtained a metallicity of [Fe/H] = -0.5. We also found that this star is a retrograde halo star with a very high eccentricity of  $0.917_{-0.025}^{+0.005}$. We obtained an age of $4.51\pm1.44$ Gyr, and a mass of 
$1.09\pm0.10$ $M_\odot$. We also found that the star likely last passed $550\;\mathrm{pc}$ away from the Galactic centre. This does not support the idea that the star was ejected from the supermassive black hole at the centre of the Galaxy.

In terms of chemical abundances, the star has a chemical pattern compatible with typical  retrograde stars, supporting the idea that the star is part of the retrograde halo. We also found that the low metallicity, combined with the high abundance of $\alpha$ elements, is an indicator that the star is old and formed in an environment with a high star formation rate. Considering that the star is part of the retrograde halo, it is possible that it was born in another galaxy and later accreted onto the Milky Way. We found an overabundance of europium  of [Eu/Fe] = 0.93 $\pm$ 0.24, indicating that J01020100-7122208 might come from the Gaia-Enceladus galaxy. 

Our age of 4~Gyr is slightly inconsistent  with our results from the chemistry, which points towards a star formed before 4~Gyr ago. We attribute this inconsistency to the difficulties regarding the age determination of metal-poor red giant stars
and also to the idea that this star could be the product of a merge between two stars, rejuvenating the star like a blue straggler.

Based on the kinematics, ages and chemical abundances of 16 elements, we concluded that this star is not likely to have been ejected from the central black hole of our Galaxy, but instead is an accreted star, probably from Gaia-Enceladus. To arrive at this conclusion, it was necessary to combine astrometric, photometric and spectroscopic information. We have demonstrated that to truly understand where a star comes from, the best is to combine information about kinematics, ages and chemical abundances.  With kinematics we can retrieve the trajectory a star has followed. With ages we can both constrain the orbits of stars and associate an object to a stellar population.  With chemical abundances we can both validate possible origins retrieved from kinematics and also point out new origins. We are entering a revolutionary time in Galactic archaeology in which we have all this information for millions of stars. 

\section*{Acknowledgements}

D.B-S. acknowledges Marcelo Tucci Maia and Claudia Aguilera-G\'omez for all the discussions and for the interest, as well as Anya Samadi and Francisca Espinoza for the support. D.B-S also acknowledges ECOS-Anid grant 140089.

The work was financially supported by the ECOS-Anid Grant 180049.  P.J. acknowledges financial support from FONDECYT Iniciaci\'on Grant Number 11170174 and Regular grant Number 1200703.

KH has been partially supported by a TDA/Scialog (2018-2020) grant funded by the Research Corporation, a TDA/Scialog grant (2019-2021) funded by the Heising-Simons Foundation, and through the Wootton Center for Astrophysical Plasma Properties funded under the United States Department of Energy collaborative agreement DE-NA0003843. KH also acknowledges support from the National Science Foundation grant AST-1907417.

Support for JLP is provided in part by ANID through the Fondecyt regular grant 1191038 and through the Millennium Science Initiative grant ICN12\_009, awarded to The Millennium Institute of Astrophysics, MAS.

This paper includes data gathered with the 6.5 meter Magellan Telescopes located at Las Campanas Observatory, Chile.

This work has made use of data from the European Space Agency (ESA) mission
{\it Gaia} (\url{https://www.cosmos.esa.int/gaia}), processed by the {\it Gaia}
Data Processing and Analysis Consortium (DPAC,
\url{https://www.cosmos.esa.int/web/gaia/dpac/consortium}). Funding for the DPAC
has been provided by national institutions, in particular the institutions
participating in the {\it Gaia} Multilateral Agreement.

This work made use of the Third Data Release of the GALAH Survey (Buder et al 2021). The GALAH Survey is based on data acquired through the Australian Astronomical Observatory, under programs: A/2013B/13 (The GALAH pilot survey); A/2014A/25, A/2015A/19, A2017A/18 (The GALAH survey phase 1); A2018A/18 (Open clusters with HERMES); A2019A/1 (Hierarchical star formation in Ori OB1); A2019A/15 (The GALAH survey phase 2); A/2015B/19, A/2016A/22, A/2016B/10, A/2017B/16, A/2018B/15 (The HERMES-TESS program); and A/2015A/3, A/2015B/1, A/2015B/19, A/2016A/22, A/2016B/12, A/2017A/14 (The HERMES K2-follow-up program). We acknowledge the traditional owners of the land on which the AAT stands, the Gamilaraay people, and pay our respects to elders past and present. This paper includes data that has been provided by AAO Data Central (datacentral.aao.gov.au).

\section*{Data Availability}

The data underlying this article will be shared on reasonable request to the corresponding author. The GALAH DR3 data used in this work can found at \url{https://www.galah-survey.org/dr3/the_catalogues/}. The GAIA EDR3 data used in this work can be found at \url{https://www.cosmos.esa.int/web/gaia/earlydr3}.

\bibliographystyle{mnras}
\bibliography{mnras_template_ysg_paper.bbl}

\begin{thebibliography}{}
\makeatletter
\relax
\def\mn@urlcharsother{\let\do\@makeother \do\$\do\&\do\#\do\^\do\_\do\%\do\~}
\def\mn@doi{\begingroup\mn@urlcharsother \@ifnextchar [ {\mn@doi@}
  {\mn@doi@[]}}
\def\mn@doi@[#1]#2{\def\@tempa{#1}\ifx\@tempa\@empty \href
  {http://dx.doi.org/#2} {doi:#2}\else \href {http://dx.doi.org/#2} {#1}\fi
  \endgroup}
\def\mn@eprint#1#2{\mn@eprint@#1:#2::\@nil}
\def\mn@eprint@arXiv#1{\href {http://arxiv.org/abs/#1} {{\tt arXiv:#1}}}
\def\mn@eprint@dblp#1{\href {http://dblp.uni-trier.de/rec/bibtex/#1.xml}
  {dblp:#1}}
\def\mn@eprint@#1:#2:#3:#4\@nil{\def\@tempa {#1}\def\@tempb {#2}\def\@tempc
  {#3}\ifx \@tempc \@empty \let \@tempc \@tempb \let \@tempb \@tempa \fi \ifx
  \@tempb \@empty \def\@tempb {arXiv}\fi \@ifundefined
  {mn@eprint@\@tempb}{\@tempb:\@tempc}{\expandafter \expandafter \csname
  mn@eprint@\@tempb\endcsname \expandafter{\@tempc}}}

\bibitem[\protect\citeauthoryear{Abolfathi et~al.,}{Abolfathi
  et~al.}{2018}]{Abolfathi2018}
Abolfathi B.,  et~al., 2018, The Astrophysical Journal Supplement Series, 235,
  42

\bibitem[\protect\citeauthoryear{{Aguado} et~al.,}{{Aguado}
  et~al.}{2021}]{aguado20}
{Aguado} D.~S.,  et~al., 2021, \mn@doi [\apjl] {10.3847/2041-8213/abdbb8},
  \href {https://ui.adsabs.harvard.edu/abs/2021ApJ...908L...8A} {908, L8}

\bibitem[\protect\citeauthoryear{Alvarez \& Plez}{Alvarez \&
  Plez}{1997}]{alvarez1997near}
Alvarez R.,  Plez B.,  1997, arXiv preprint astro-ph/9710157

\bibitem[\protect\citeauthoryear{Amarsi, Nordlander, Barklem, Asplund, Collet
  \& Lind}{Amarsi et~al.}{2018}]{amarsi2018effective}
Amarsi A.~M.,  Nordlander T.,  Barklem P.,  Asplund M.,  Collet R.,   Lind K.,
  2018, Astronomy \& Astrophysics, 615, A139

\bibitem[\protect\citeauthoryear{Anders et~al.,}{Anders
  et~al.}{2017}]{anders2017galactic}
Anders F.,  et~al., 2017, Astronomy \& Astrophysics, 597, A30

\bibitem[\protect\citeauthoryear{Baglin et~al.,}{Baglin
  et~al.}{2006}]{baglin2006scientific}
Baglin A.,  et~al., 2006, in The CoRoT Mission Pre-Launch Status-Stellar
  Seismology and Planet Finding. p.~33

\bibitem[\protect\citeauthoryear{Belokurov, Erkal, Evans, Koposov  \&
  Deason}{Belokurov et~al.}{2018}]{Belokurov2018}
Belokurov V.,  Erkal D.,  Evans N.,  Koposov S.,   Deason A.,  2018, Monthly
  Notices of the Royal Astronomical Society, 478, 611

\bibitem[\protect\citeauthoryear{{Bennett} \& {Bovy}}{{Bennett} \&
  {Bovy}}{2019}]{bennett2019}
{Bennett} M.,  {Bovy} J.,  2019, \mn@doi [\mnras] {10.1093/mnras/sty2813},
  \href {https://ui.adsabs.harvard.edu/abs/2019MNRAS.482.1417B} {482, 1417}

\bibitem[\protect\citeauthoryear{Bissantz, Englmaier  \& Gerhard}{Bissantz
  et~al.}{2003}]{bissantz2003gas}
Bissantz N.,  Englmaier P.,   Gerhard O.,  2003, Monthly Notices of the Royal
  Astronomical Society, 340, 949

\bibitem[\protect\citeauthoryear{Blanco-Cuaresma}{Blanco-Cuaresma}{2019}]{Blanco2019}
Blanco-Cuaresma S.,  2019, Monthly Notices of the Royal Astronomical Society,
  486, 2075

\bibitem[\protect\citeauthoryear{Blanco-Cuaresma, Soubiran, Jofr{\'e}  \&
  Heiter}{Blanco-Cuaresma et~al.}{2014a}]{blanco2014gaia}
Blanco-Cuaresma S.,  Soubiran C.,  Jofr{\'e} P.,   Heiter U.,  2014a, Astronomy
  \& Astrophysics, 566, A98

\bibitem[\protect\citeauthoryear{Blanco-Cuaresma, Soubiran, Heiter  \&
  Jofr{\'e}}{Blanco-Cuaresma et~al.}{2014b}]{Blanco2014}
Blanco-Cuaresma S.,  Soubiran C.,  Heiter U.,   Jofr{\'e} P.,  2014b, Astronomy
  \& Astrophysics, 569, A111

\bibitem[\protect\citeauthoryear{Borucki et~al.,}{Borucki
  et~al.}{2010}]{borucki2010kepler}
Borucki W.~J.,  et~al., 2010, Science, 327, 977

\bibitem[\protect\citeauthoryear{Brown et~al.,}{Brown
  et~al.}{2018}]{brown2018gaia}
Brown A.,  et~al., 2018, Astronomy \& astrophysics, 616, A1

\bibitem[\protect\citeauthoryear{Brown et~al.,}{Brown
  et~al.}{2021}]{brown2021gaia}
Brown A.~G.,  et~al., 2021, Astronomy \& Astrophysics, 649, A1

\bibitem[\protect\citeauthoryear{Buchner et~al.,}{Buchner
  et~al.}{2014}]{buchner2014x}
Buchner J.,  et~al., 2014, Astronomy \& Astrophysics, 564, A125

\bibitem[\protect\citeauthoryear{Buder et~al.,}{Buder
  et~al.}{2020}]{buder2020galah+}
Buder S.,  et~al., 2020, Monthly Notices of the Royal Astronomical Society

\bibitem[\protect\citeauthoryear{Busso, Gallino  \& Wasserburg}{Busso
  et~al.}{1999}]{busso1999nucleosynthesis}
Busso M.,  Gallino R.,   Wasserburg G.,  1999, Annual Review of Astronomy and
  Astrophysics, 37, 239

\bibitem[\protect\citeauthoryear{Carollo et~al.,}{Carollo
  et~al.}{2007}]{carollo2007two}
Carollo D.,  et~al., 2007, Nature, 450, 1020

\bibitem[\protect\citeauthoryear{Casamiquela, Tarricq, Soubiran,
  Blanco-Cuaresma, Jofr{\'e}, Heiter  \& Maia}{Casamiquela
  et~al.}{2019}]{casamiquela2019}
Casamiquela L.,  Tarricq Y.,  Soubiran C.,  Blanco-Cuaresma S.,  Jofr{\'e} P.,
  Heiter U.,   Maia M.~T.,  2019, arXiv preprint arXiv:1912.08539

\bibitem[\protect\citeauthoryear{Chiappini et~al.,}{Chiappini
  et~al.}{2015}]{chiappini2015young}
Chiappini C.,  et~al., 2015, Astronomy \& Astrophysics, 576, L12

\bibitem[\protect\citeauthoryear{Choi, Dotter, Conroy, Cantiello, Paxton  \&
  Johnson}{Choi et~al.}{2016}]{choi2016mesa}
Choi J.,  Dotter A.,  Conroy C.,  Cantiello M.,  Paxton B.,   Johnson B.~D.,
  2016, The Astrophysical Journal, 823, 102

\bibitem[\protect\citeauthoryear{Combes}{Combes}{2013}]{combes2013gas}
Combes F.,  2013, arXiv preprint arXiv:1309.1603

\bibitem[\protect\citeauthoryear{Das, Hawkins  \& Jofr{\'e}}{Das
  et~al.}{2020}]{das2020ages}
Das P.,  Hawkins K.,   Jofr{\'e} P.,  2020, Monthly Notices of the Royal
  Astronomical Society, 493, 5195

\bibitem[\protect\citeauthoryear{Dotter}{Dotter}{2016}]{dotter2016mesa}
Dotter A.,  2016, The Astrophysical Journal Supplement Series, 222, 8

\bibitem[\protect\citeauthoryear{{Drimmel} \& {Poggio}}{{Drimmel} \&
  {Poggio}}{2018}]{drimmel2018}
{Drimmel} R.,  {Poggio} E.,  2018, \mn@doi [Research Notes of the American
  Astronomical Society] {10.3847/2515-5172/aaef8b}, \href
  {https://ui.adsabs.harvard.edu/abs/2018RNAAS...2..210D} {2, 210}

\bibitem[\protect\citeauthoryear{{Eisenhardt} et~al.,}{{Eisenhardt}
  et~al.}{2020}]{eisenhardt2020}
{Eisenhardt} P. R.~M.,  et~al., 2020, \mn@doi [\apjs]
  {10.3847/1538-4365/ab7f2a}, \href
  {https://ui.adsabs.harvard.edu/abs/2020ApJS..247...69E} {247, 69}

\bibitem[\protect\citeauthoryear{Epstein et~al.,}{Epstein
  et~al.}{2014}]{epstein2014testing}
Epstein C.~R.,  et~al., 2014, The Astrophysical Journal Letters, 785, L28

\bibitem[\protect\citeauthoryear{Feroz \& Hobson}{Feroz \&
  Hobson}{2008}]{feroz2008multimodal}
Feroz F.,  Hobson M.~P.,  2008, Monthly Notices of the Royal Astronomical
  Society, 384, 449

\bibitem[\protect\citeauthoryear{Feuillet, Feltzing, Sahlholdt  \&
  Casagrande}{Feuillet et~al.}{2020}]{feuillet2020skymapper}
Feuillet D.~K.,  Feltzing S.,  Sahlholdt C.~L.,   Casagrande L.,  2020, Monthly
  Notices of the Royal Astronomical Society, 497, 109

\bibitem[\protect\citeauthoryear{{Gaia Collaboration} et~al.,}{{Gaia
  Collaboration} et~al.}{2016}]{GaiaCollaboration+2016b}
{Gaia Collaboration} et~al., 2016, \mn@doi [\aap]
  {10.1051/0004-6361/201629272}, \href
  {http://adsabs.harvard.edu/abs/2016A%26A...595A...1G} {595, A1}

\bibitem[\protect\citeauthoryear{Gehren}{Gehren}{1981}]{gehren1981temperature}
Gehren T.,  1981, Astronomy and Astrophysics, 100, 97

\bibitem[\protect\citeauthoryear{{Gilmore} \& {Wyse}}{{Gilmore} \&
  {Wyse}}{1998}]{gilmorewyse98}
{Gilmore} G.,  {Wyse} R. F.~G.,  1998, \mn@doi [\aj] {10.1086/300459}, \href
  {https://ui.adsabs.harvard.edu/abs/1998AJ....116..748G} {116, 748}

\bibitem[\protect\citeauthoryear{Gratton, Carretta, Desidera, Lucatello, Mazzei
   \& Barbieri}{Gratton et~al.}{2003}]{gratton2003abundances}
Gratton R.,  Carretta E.,  Desidera S.,  Lucatello S.,  Mazzei P.,   Barbieri
  M.,  2003, Astronomy \& Astrophysics, 406, 131

\bibitem[\protect\citeauthoryear{{Gravity Collaboration} et~al.,}{{Gravity
  Collaboration} et~al.}{2018}]{gravity2018}
{Gravity Collaboration} et~al., 2018, \mn@doi [\aap]
  {10.1051/0004-6361/201833718}, \href
  {https://ui.adsabs.harvard.edu/abs/2018A&A...615L..15G} {615, L15}

\bibitem[\protect\citeauthoryear{Gray}{Gray}{2005}]{gray2005observation}
Gray D.~F.,  2005, The observation and analysis of stellar photospheres.
Cambridge University Press

\bibitem[\protect\citeauthoryear{Grevesse, Asplund  \& Sauval}{Grevesse
  et~al.}{2007}]{grevesse2007}
Grevesse N.,  Asplund M.,   Sauval A.,  2007, Space Science Reviews, 130, 105

\bibitem[\protect\citeauthoryear{Gustafsson, Edvardsson, Eriksson,
  J{\o}rgensen, Nordlund  \& Plez}{Gustafsson et~al.}{2008}]{gustafsson2008}
Gustafsson B.,  Edvardsson B.,  Eriksson K.,  J{\o}rgensen U.~G.,  Nordlund
  {\AA}.,   Plez B.,  2008, Astronomy \& Astrophysics, 486, 951

\bibitem[\protect\citeauthoryear{Hansen, Rich, Koch, Xu, Kunder  \&
  Ludwig}{Hansen et~al.}{2016}]{hansen2016chemical}
Hansen C.~J.,  Rich R.~M.,  Koch A.,  Xu S.,  Kunder A.,   Ludwig H.-G.,  2016,
  Astronomy \& Astrophysics, 590, A39

\bibitem[\protect\citeauthoryear{Hawkins \& Wyse}{Hawkins \&
  Wyse}{2018}]{hawkins2018fastest}
Hawkins K.,  Wyse R.~F.,  2018, Monthly Notices of the Royal Astronomical
  Society, 481, 1028

\bibitem[\protect\citeauthoryear{Hawkins, Jofre, Gilmore  \& Masseron}{Hawkins
  et~al.}{2014}]{hawkins2014relative}
Hawkins K.,  Jofre P.,  Gilmore G.,   Masseron T.,  2014, Monthly Notices of
  the Royal Astronomical Society, 445, 2575

\bibitem[\protect\citeauthoryear{Hawkins et~al.,}{Hawkins
  et~al.}{2016}]{hawkins2016gaia}
Hawkins K.,  et~al., 2016, Astronomy \& Astrophysics, 592, A70

\bibitem[\protect\citeauthoryear{Heiter et~al.,}{Heiter
  et~al.}{2015a}]{heiter2015atomic}
Heiter U.,  et~al., 2015a, Physica Scripta, 90, 054010

\bibitem[\protect\citeauthoryear{Heiter, Jofr{\'e}, Gustafsson, Korn, Soubiran
  \& Th{\'e}venin}{Heiter et~al.}{2015b}]{heiter2015benchmark}
Heiter U.,  Jofr{\'e} P.,  Gustafsson B.,  Korn A.~J.,  Soubiran C.,
  Th{\'e}venin F.,  2015b, Astronomy \& Astrophysics, 582, A49

\bibitem[\protect\citeauthoryear{Hekker \& Johnson}{Hekker \&
  Johnson}{2019}]{hekker2019origin}
Hekker S.,  Johnson J.~A.,  2019, Monthly Notices of the Royal Astronomical
  Society, 487, 4343

\bibitem[\protect\citeauthoryear{Helmi, Babusiaux, Koppelman, Massari,
  Veljanoski  \& Brown}{Helmi et~al.}{2018}]{Helmi2018}
Helmi A.,  Babusiaux C.,  Koppelman H.~H.,  Massari D.,  Veljanoski J.,   Brown
  A.~G.,  2018, Nature, 563, 85

\bibitem[\protect\citeauthoryear{Holtzman et~al.,}{Holtzman
  et~al.}{2018}]{holtzman2018apogee}
Holtzman J.~A.,  et~al., 2018, The Astronomical Journal, 156, 125

\bibitem[\protect\citeauthoryear{{Iwamoto}, {Brachwitz}, {Nomoto}, {Kishimoto},
  {Umeda}, {Hix}  \& {Thielemann}}{{Iwamoto} et~al.}{1999}]{iwamoto99}
{Iwamoto} K.,  {Brachwitz} F.,  {Nomoto} K.,  {Kishimoto} N.,  {Umeda} H.,
  {Hix} W.~R.,   {Thielemann} F.-K.,  1999, \mn@doi [\apjs] {10.1086/313278},
  \href {https://ui.adsabs.harvard.edu/abs/1999ApJS..125..439I} {125, 439}

\bibitem[\protect\citeauthoryear{Izzard, Preece, Jofre, Halabi, Masseron  \&
  Tout}{Izzard et~al.}{2018}]{izzard2018binary}
Izzard R.~G.,  Preece H.,  Jofre P.,  Halabi G.~M.,  Masseron T.,   Tout C.~A.,
   2018, Monthly Notices of the Royal Astronomical Society, 473, 2984

\bibitem[\protect\citeauthoryear{Jofre \& Weiss}{Jofre \&
  Weiss}{2011}]{jofre2011age}
Jofre P.,  Weiss A.,  2011, Astronomy \& Astrophysics, 533, A59

\bibitem[\protect\citeauthoryear{Jofr{\'e} et~al.,}{Jofr{\'e}
  et~al.}{2014}]{jofre2014gaia}
Jofr{\'e} P.,  et~al., 2014, Astronomy \& astrophysics, 564, A133

\bibitem[\protect\citeauthoryear{Jofr{\'e} et~al.,}{Jofr{\'e}
  et~al.}{2015}]{jofre2015gaia}
Jofr{\'e} P.,  et~al., 2015, Astronomy \& astrophysics, 582, A81

\bibitem[\protect\citeauthoryear{Jofr{\'e} et~al.,}{Jofr{\'e}
  et~al.}{2016}]{jofre2016cannibals}
Jofr{\'e} P.,  et~al., 2016, Astronomy \& Astrophysics, 595, A60

\bibitem[\protect\citeauthoryear{Jofr{\'e} et~al.,}{Jofr{\'e}
  et~al.}{2017}]{jofre2017gaia}
Jofr{\'e} P.,  et~al., 2017, Astronomy \& Astrophysics, 601, A38

\bibitem[\protect\citeauthoryear{Jofr{\'e}, Heiter  \& Soubiran}{Jofr{\'e}
  et~al.}{2019}]{Jofre2019}
Jofr{\'e} P.,  Heiter U.,   Soubiran C.,  2019, Annual Review of Astronomy and
  Astrophysics, 57, 571

\bibitem[\protect\citeauthoryear{Juri{\'c} et~al.,}{Juri{\'c}
  et~al.}{2008}]{juric2008milky}
Juri{\'c} M.,  et~al., 2008, The Astrophysical Journal, 673, 864

\bibitem[\protect\citeauthoryear{Kalirai}{Kalirai}{2012}]{kalirai2012age}
Kalirai J.~S.,  2012, Nature, 486, 90

\bibitem[\protect\citeauthoryear{Kelson}{Kelson}{2003}]{kelson2003optimal}
Kelson D.~D.,  2003, Publications of the Astronomical Society of the Pacific,
  115, 688

\bibitem[\protect\citeauthoryear{Kobayashi \& Nomoto}{Kobayashi \&
  Nomoto}{2009}]{kobayashi2009role}
Kobayashi C.,  Nomoto K.,  2009, The Astrophysical Journal, 707, 1466

\bibitem[\protect\citeauthoryear{Kobayashi, Umeda, Nomoto, Tominaga  \&
  Ohkubo}{Kobayashi et~al.}{2006}]{kobayashi2006galactic}
Kobayashi C.,  Umeda H.,  Nomoto K.,  Tominaga N.,   Ohkubo T.,  2006, The
  Astrophysical Journal, 653, 1145

\bibitem[\protect\citeauthoryear{Koposov}{Koposov}{2020}]{sergey_koposov_2020_4002972}
Koposov S.,  2020, \mn@doi [Zenodo] {10.5281/zenodo.4002972}

\bibitem[\protect\citeauthoryear{{Koposov} et~al.,}{{Koposov}
  et~al.}{2020}]{koposov2020}
{Koposov} S.~E.,  et~al., 2020, \mn@doi [\mnras] {10.1093/mnras/stz3081}, \href
  {https://ui.adsabs.harvard.edu/abs/2020MNRAS.491.2465K} {491, 2465}

\bibitem[\protect\citeauthoryear{Koppelman, Helmi, Massari, Price-Whelan  \&
  Starkenburg}{Koppelman et~al.}{2019}]{Koppelman2019}
Koppelman H.~H.,  Helmi A.,  Massari D.,  Price-Whelan A.~M.,   Starkenburg
  T.~K.,  2019, Astronomy \& Astrophysics, 631, L9

\bibitem[\protect\citeauthoryear{Koppelman, Bos  \& Helmi}{Koppelman
  et~al.}{2020}]{koppelman2020massive}
Koppelman H.~H.,  Bos R.~O.,   Helmi A.,  2020, arXiv preprint arXiv:2006.07620

\bibitem[\protect\citeauthoryear{Kordopatis, Recio-Blanco, Schultheis  \&
  Hill}{Kordopatis et~al.}{2020}]{kordopatis2020chemodynamics}
Kordopatis G.,  Recio-Blanco A.,  Schultheis M.,   Hill V.,  2020, Astronomy \&
  Astrophysics, 643, A69

\bibitem[\protect\citeauthoryear{{Lindegren} et~al.,}{{Lindegren}
  et~al.}{2021}]{lindegren2021}
{Lindegren} L.,  et~al., 2021, \mn@doi [\aap] {10.1051/0004-6361/202039709},
  \href {https://ui.adsabs.harvard.edu/abs/2021A&A...649A...2L} {649, A2}

\bibitem[\protect\citeauthoryear{Mackereth et~al.,}{Mackereth
  et~al.}{2018}]{mackereth2018origin}
Mackereth J.~T.,  et~al., 2018, arXiv preprint arXiv:1808.00968

\bibitem[\protect\citeauthoryear{Martig et~al.,}{Martig
  et~al.}{2015}]{martig2015young}
Martig M.,  et~al., 2015, Monthly Notices of the Royal Astronomical Society,
  451, 2230

\bibitem[\protect\citeauthoryear{Massey, Levine, Neugent, Levesque, Morrell  \&
  Skiff}{Massey et~al.}{2018}]{massey2018runaway}
Massey P.,  Levine S.~E.,  Neugent K.~F.,  Levesque E.,  Morrell N.,   Skiff
  B.,  2018, The Astronomical Journal, 156, 265

\bibitem[\protect\citeauthoryear{Matsuno, Yong, Aoki  \& Ishigaki}{Matsuno
  et~al.}{2018}]{matsuno2018optical}
Matsuno T.,  Yong D.,  Aoki W.,   Ishigaki M.~N.,  2018, The Astrophysical
  Journal, 860, 49

\bibitem[\protect\citeauthoryear{Matsuno, Aoki  \& Suda}{Matsuno
  et~al.}{2019}]{matsuno2019origin}
Matsuno T.,  Aoki W.,   Suda T.,  2019, The Astrophysical Journal Letters, 874,
  L35

\bibitem[\protect\citeauthoryear{Matsuno et~al.,}{Matsuno
  et~al.}{2020}]{matsuno2020star}
Matsuno T.,  et~al., 2020, arXiv preprint arXiv:2006.03619

\bibitem[\protect\citeauthoryear{{Matsuno}, {Hirai}, {Tarumi}, {Hotokezaka},
  {Tanaka}  \& {Helmi}}{{Matsuno} et~al.}{2021}]{matsuno21}
{Matsuno} T.,  {Hirai} Y.,  {Tarumi} Y.,  {Hotokezaka} K.,  {Tanaka} M.,
  {Helmi} A.,  2021, arXiv e-prints, \href
  {https://ui.adsabs.harvard.edu/abs/2021arXiv210107791M} {p. arXiv:2101.07791}

\bibitem[\protect\citeauthoryear{Matteucci \& Greggio}{Matteucci \&
  Greggio}{1986}]{matteucci1986relative}
Matteucci F.,  Greggio L.,  1986, Astronomy and Astrophysics, 154, 279

\bibitem[\protect\citeauthoryear{Matteucci, Romano, Arcones, Korobkin  \&
  Rosswog}{Matteucci et~al.}{2014}]{matteucci2014europium}
Matteucci F.,  Romano D.,  Arcones A.,  Korobkin O.,   Rosswog S.,  2014,
  \mn@doi [Monthly Notices of the Royal Astronomical Society]
  {10.1093/mnras/stt2350}, 438, 2177

\bibitem[\protect\citeauthoryear{McMillan}{McMillan}{2016}]{mcmillan2016mass}
McMillan P.~J.,  2016, Monthly Notices of the Royal Astronomical Society, p.
  stw2759

\bibitem[\protect\citeauthoryear{Montalb{\'a}n et~al.,}{Montalb{\'a}n
  et~al.}{2021}]{montalban2021chronologically}
Montalb{\'a}n J.,  et~al., 2021, Nature Astronomy, pp~1--8

\bibitem[\protect\citeauthoryear{Myeong, Vasiliev, Iorio, Evans  \&
  Belokurov}{Myeong et~al.}{2019}]{myeong2019evidence}
Myeong G.,  Vasiliev E.,  Iorio G.,  Evans N.,   Belokurov V.,  2019, Monthly
  Notices of the Royal Astronomical Society, 488, 1235

\bibitem[\protect\citeauthoryear{Neugent, Massey, Skiff, Drout, Meynet  \&
  Olsen}{Neugent et~al.}{2010}]{neugent2010yellow}
Neugent K.~F.,  Massey P.,  Skiff B.,  Drout M.~R.,  Meynet G.,   Olsen K.~A.,
  2010, The Astrophysical Journal, 719, 1784

\bibitem[\protect\citeauthoryear{Neugent, Massey, Morrell, Skiff  \&
  Georgy}{Neugent et~al.}{2018}]{neugent2018runaway}
Neugent K.~F.,  Massey P.,  Morrell N.~I.,  Skiff B.,   Georgy C.,  2018, The
  Astronomical Journal, 155, 207

\bibitem[\protect\citeauthoryear{Nissen \& Schuster}{Nissen \&
  Schuster}{2010}]{nissen2010two}
Nissen P.~E.,  Schuster W.~J.,  2010, Astronomy \& Astrophysics, 511, L10

\bibitem[\protect\citeauthoryear{Nomoto, Kobayashi  \& Tominaga}{Nomoto
  et~al.}{2013}]{nomoto2013nucleosynthesis}
Nomoto K.,  Kobayashi C.,   Tominaga N.,  2013, Annual Review of Astronomy and
  Astrophysics, 51, 457

\bibitem[\protect\citeauthoryear{{Onken} et~al.,}{{Onken}
  et~al.}{2019}]{onken2019}
{Onken} C.~A.,  et~al., 2019, \mn@doi [\pasa] {10.1017/pasa.2019.27}, \href
  {https://ui.adsabs.harvard.edu/abs/2019PASA...36...33O} {36, e033}

\bibitem[\protect\citeauthoryear{Piffl et~al.,}{Piffl
  et~al.}{2014}]{piffl2014rave}
Piffl T.,  et~al., 2014, Astronomy \& Astrophysics, 562, A91

\bibitem[\protect\citeauthoryear{Pignatari, Gallino, Heil, Wiescher,
  K{\"a}ppeler, Herwig  \& Bisterzo}{Pignatari
  et~al.}{2010}]{pignatari2010weak}
Pignatari M.,  Gallino R.,  Heil M.,  Wiescher M.,  K{\"a}ppeler F.,  Herwig
  F.,   Bisterzo S.,  2010, The Astrophysical Journal, 710, 1557

\bibitem[\protect\citeauthoryear{Plez}{Plez}{2012}]{plez2012turbospectrum}
Plez B.,  2012, Astrophysics Source Code Library, pp ascl--1205

\bibitem[\protect\citeauthoryear{Price-Whelan}{Price-Whelan}{2017}]{gala}
Price-Whelan A.~M.,  2017, \mn@doi [The Journal of Open Source Software]
  {10.21105/joss.00388}, 2

\bibitem[\protect\citeauthoryear{Price-Whelan et~al.,}{Price-Whelan
  et~al.}{2018}]{price2018astropy}
Price-Whelan A.~M.,  et~al., 2018, The Astronomical Journal, 156, 123

\bibitem[\protect\citeauthoryear{Randich \& Magrini}{Randich \&
  Magrini}{2021}]{randich2021light}
Randich S.,  Magrini L.,  2021, Frontiers in Astronomy and Space Sciences, 8, 6

\bibitem[\protect\citeauthoryear{{Reid} \& {Brunthaler}}{{Reid} \&
  {Brunthaler}}{2004}]{reid2004}
{Reid} M.~J.,  {Brunthaler} A.,  2004, \mn@doi [\apj] {10.1086/424960}, \href
  {https://ui.adsabs.harvard.edu/abs/2004ApJ...616..872R} {616, 872}

\bibitem[\protect\citeauthoryear{{Riello} et~al.,}{{Riello}
  et~al.}{2021}]{riello2021}
{Riello} M.,  et~al., 2021, \mn@doi [\aap] {10.1051/0004-6361/202039587}, \href
  {https://ui.adsabs.harvard.edu/abs/2021A&A...649A...3R} {649, A3}

\bibitem[\protect\citeauthoryear{Robitaille et~al.,}{Robitaille
  et~al.}{2013}]{robitaille2013astropy}
Robitaille T.~P.,  et~al., 2013, Astronomy \& Astrophysics, 558, A33

\bibitem[\protect\citeauthoryear{Rossi, Marchetti, Cacciato, Kuiack  \&
  Sari}{Rossi et~al.}{2017}]{rossi2017joint}
Rossi E.~M.,  Marchetti T.,  Cacciato M.,  Kuiack M.,   Sari R.,  2017, Monthly
  Notices of the Royal Astronomical Society, 467, 1844

\bibitem[\protect\citeauthoryear{Ruchti, Bergemann, Serenelli, Casagrande  \&
  Lind}{Ruchti et~al.}{2013}]{ruchti2013unveiling}
Ruchti G.~R.,  Bergemann M.,  Serenelli A.,  Casagrande L.,   Lind K.,  2013,
  Monthly Notices of the Royal Astronomical Society, 429, 126

\bibitem[\protect\citeauthoryear{Schuster, Moreno, Nissen  \&
  Pichardo}{Schuster et~al.}{2012}]{schuster2012two}
Schuster W.~J.,  Moreno E.,  Nissen P.~E.,   Pichardo B.,  2012, Astronomy \&
  Astrophysics, 538, A21

\bibitem[\protect\citeauthoryear{Searle \& Oke}{Searle \&
  Oke}{1962}]{searle1962effective}
Searle L.,  Oke J.,  1962, The Astrophysical Journal, 135, 790

\bibitem[\protect\citeauthoryear{Silva~Aguirre et~al.,}{Silva~Aguirre
  et~al.}{2018}]{aguirre2018confirming}
Silva~Aguirre V.,  et~al., 2018, Monthly Notices of the Royal Astronomical
  Society

\bibitem[\protect\citeauthoryear{{Skrutskie} et~al.,}{{Skrutskie}
  et~al.}{2006}]{skrutskie2006}
{Skrutskie} M.~F.,  et~al., 2006, \mn@doi [\aj] {10.1086/498708}, \href
  {https://ui.adsabs.harvard.edu/abs/2006AJ....131.1163S} {131, 1163}

\bibitem[\protect\citeauthoryear{{Tayar} et~al.,}{{Tayar}
  et~al.}{2017}]{tayar2017}
{Tayar} J.,  et~al., 2017, \mn@doi [\apj] {10.3847/1538-4357/aa6a1e}, \href
  {https://ui.adsabs.harvard.edu/abs/2017ApJ...840...17T} {840, 17}

\bibitem[\protect\citeauthoryear{Timmes, Woosley  \& Weaver}{Timmes
  et~al.}{1995}]{timmes1995galactic}
Timmes F.,  Woosley S.,   Weaver T.~A.,  1995, The Astrophysical Journal
  Supplement Series, 98, 617

\bibitem[\protect\citeauthoryear{Tinsley}{Tinsley}{1979}]{tinsley1979stellar}
Tinsley B.,  1979, The Astrophysical Journal, 229, 1046

\bibitem[\protect\citeauthoryear{{Tody}}{{Tody}}{1993}]{iraf1993}
{Tody} D.,  1993, in {Hanisch} R.~J.,  {Brissenden} R.~J.~V.,   {Barnes} J.,
  eds,  Astronomical Society of the Pacific Conference Series Vol. 52,
  Astronomical Data Analysis Software and Systems II. p.~173

\bibitem[\protect\citeauthoryear{Tolstoy, Hill  \& Tosi}{Tolstoy
  et~al.}{2009}]{tolstoy2009star}
Tolstoy E.,  Hill V.,   Tosi M.,  2009, Annual Review of Astronomy and
  Astrophysics, 47, 371

\bibitem[\protect\citeauthoryear{Vincenzo, Spitoni, Calura, Matteucci,
  Silva~Aguirre, Miglio  \& Cescutti}{Vincenzo et~al.}{2019}]{vincenzo2019fall}
Vincenzo F.,  Spitoni E.,  Calura F.,  Matteucci F.,  Silva~Aguirre V.,  Miglio
  A.,   Cescutti G.,  2019, Monthly Notices of the Royal Astronomical Society:
  Letters, 487, L47

\bibitem[\protect\citeauthoryear{Woosley \& Weaver}{Woosley \&
  Weaver}{1995}]{woosley1995evolution}
Woosley S.,  Weaver T.~A.,  1995, Technical report, The evolution and explosion
  of massive Stars II: Explosive hydrodynamics and nucleosynthesis.
Lawrence Livermore National Lab., CA (United States)

\bibitem[\protect\citeauthoryear{Yong et~al.,}{Yong
  et~al.}{2016}]{yong2016graces}
Yong D.,  et~al., 2016, Monthly Notices of the Royal Astronomical Society, 459,
  487

\bibitem[\protect\citeauthoryear{Zhang et~al.,}{Zhang
  et~al.}{2021}]{zhang2021high}
Zhang H.,  et~al., 2021, arXiv preprint arXiv:2102.05999

\makeatother
\end{thebibliography}

\appendix

\section{Adopted line regions}\label{line_regions}

In order to calculate the stellar parameters of J01020100-7122208, we used a list of lines adjusted on the list \textit{synthe\_synth\_good\_for\_params}, which is provided  with the package of \texttt{iSpec}. This list was built using as a basis the Gaia-ESO linelist.

 With the aim of having a better agreement between the values of stellar parameters calculated for our control sample and the reference values provided by APOGEE survey, we considered regions sensitive to stellar parameters, in particular those with Fe 2 lines that help with the calculation of the surface gravity. In Table \ref{tab:regions_stellarparam}, we present the lines used when calculating stellar parameters.
The lines used to calculate chemical abundances are presented in Table \ref{tab:regions_chemical}.

\begin{table*}
\caption{Lines used to calculate stellar parameters of J01020100-7122208.}
\label{tab:regions_stellarparam}
\begin{tabular}{|r|l|r|l|r|l|r|l|r|l|r|l|}
\hline
  \multicolumn{1}{|c|}{Wavelength} &
  \multicolumn{1}{c|}{Element} &
  \multicolumn{1}{c|}{Wavelength} &
  \multicolumn{1}{c|}{Element} &
  \multicolumn{1}{c|}{Wavelength} &
  \multicolumn{1}{c|}{Element} &
  \multicolumn{1}{c|}{Wavelength} &
  \multicolumn{1}{c|}{Element} &
  \multicolumn{1}{c|}{Wavelength} &
  \multicolumn{1}{c|}{Element} &
  \multicolumn{1}{c|}{Wavelength} &
  \multicolumn{1}{c|}{Element} \\
\hline
  482.4127 & Cr 2 & 507.9740 & Fe 1 & 531.2856 & Cr 1 & 550.6779 & Fe 1 & 583.8372 & Fe 1 & 624.0310 & Fe 1\\
  482.0417 & Zr 2 & 508.3338 & Fe 1 & 531.7525 & Fe 1 & 551.2257 & Fe 1 & 584.6993 & Ni 1 & 624.0646 & Fe 1\\
  482.9373 & Cr 1 & 508.4096 & Ni 1 & 531.8771 & Cr 1 & 551.4435 & Cr 1 & 585.2293 & Fe 2 & 624.3815 & Si 1\\
  483.8556 & Fe 2 & 508.7058 & Ti 1 & 531.9035 & Fe 1 & 554.6990 & Fe 1 & 585.5076 & Fe 1 & 624.6318 & Fe 1\\
  488.1591 & Mn 1 & 509.0773 & Fe 1 & 532.0036 & Fe 1 & 554.9949 & Fe 1 & 585.7752 & Ni 1 & 625.2555 & Fe 1\\
  491.5229 & Ti 1 & 509.9930 & Ni 1 & 532.2021 & Fe 1 & 556.5541 & Fe 2 & 585.8778 & Fe 1 & 625.9595 & Ni 1\\
  491.8012 & Fe 1 & 510.4030 & Fe 1 & 532.5552 & Fe 2 & 556.7351 & Fe 1 & 585.9586 & Fe 1 & 626.5132 & Fe 1\\
  491.8994 & Fe 1 & 511.0413 & Fe 1 & 532.6161 & Fe 1 & 556.9618 & Fe 1 & 586.1109 & Fe 1 & 627.1278 & Fe 1\\
  491.9861 & Ti 1 & 511.5392 & Ni 1 & 532.7252 & Fe 1 & 557.2842 & Fe 1 & 586.7562 & Ca 1 & 630.1500 & Fe 1\\
  492.0502 & Fe 1 & 512.0415 & Ti 1 & 532.9784 & Cr 1 & 557.3102 & Fe 1 & 587.7788 & Fe 1 & 630.2493 & Fe 1\\
  493.6335 & Cr 1 & 512.4619 & Fe 1 & 532.9989 & Fe 1 & 557.6089 & Fe 1 & 589.9292 & Ti 1 & 631.5306 & Fe 1\\
  493.7348 & Ni 1 & 512.5117 & Fe 1 & 533.1481 & Fe 2 & 558.6756 & Fe 1 & 590.2473 & Fe 1 & 631.5811 & Fe 1\\
  493.8254 & Ti 1 & 512.7359 & Fe 1 & 533.2900 & Fe 1 & 558.8749 & Ca 1 & 590.3319 & Fe 2 & 631.8018 & Fe 1\\
  493.8814 & Fe 1 & 513.0359 & Ni 1 & 533.6786 & Ti 2 & 561.4773 & Ni 1 & 591.0003 & Fe 2 & 632.2166 & Ni 1\\
  494.5444 & Ni 1 & 513.2661 & Fe 2 & 533.9929 & Fe 1 & 561.5311 & Ti 2 & 590.5671 & Fe 1 & 633.5330 & Fe 1\\
  494.5636 & Fe 1 & 513.6795 & Fe 2 & 534.0447 & Cr 1 & 561.5644 & Fe 1 & 592.2110 & Ti 1 & 633.6823 & Fe 1\\
  494.6387 & Fe 1 & 516.6254 & Fe 1 & 534.8314 & Cr 1 & 561.8632 & Fe 1 & 593.0180 & Fe 1 & 633.9112 & Ni 1\\
  495.7596 & Fe 1 & 516.7954 & Cr 1 & 538.6333 & Fe 1 & 562.4542 & Fe 1 & 593.4654 & Fe 1 & 636.6481 & Ni 1\\
  496.2572 & Fe 1 & 516.9345 & Fe 1 & 538.6968 & Cr 1 & 562.8642 & Cr 1 & 594.1733 & Ti 1 & 636.9459 & Fe 2\\
  496.4927 & Cr 1 & 517.1607 & Fe 2 & 538.9479 & Fe 1 & 563.8262 & Fe 1 & 594.9346 & Fe 1 & 637.8247 & Ni 1\\
  496.6088 & Fe 1 & 517.2281 & Fe 1 & 539.2331 & Ni 1 & 564.1881 & Ni 1 & 595.2718 & Fe 1 & 640.0317 & Fe 1\\
  496.8638 & Fe 2 & 517.3186 & Fe 2 & 539.3167 & Fe 1 & 564.5613 & Si 1 & 595.3179 & Ti 1 & 641.1648 & Fe 1\\
  496.9917 & Fe 1 & 517.3782 & Ti 1 & 539.6627 & Fe 2 & 564.9699 & Ni 1 & 595.8324 & Fe 2 & 641.4581 & Ni 1\\
  497.3102 & Fe 1 & 518.3065 & Fe 1 & 539.8279 & Fe 1 & 565.5493 & Fe 1 & 596.5831 & Fe 1 & 641.4980 & Si 1\\
  497.6130 & Ni 1 & 518.4323 & Fe 1 & 540.0501 & Fe 1 & 565.8613 & Fe 2 & 597.8541 & Ti 1 & 641.9644 & Fe 1\\
  497.6325 & Ni 1 & 519.4036 & Fe 1 & 540.1340 & Ti 1 & 566.1345 & Fe 1 & 600.7960 & Fe 1 & 641.9949 & Fe 1\\
  497.7648 & Fe 1 & 521.1530 & Ti 2 & 540.3822 & Fe 1 & 566.2150 & Ti 1 & 602.7051 & Fe 1 & 642.1350 & Fe 1\\
  497.8191 & Ti 1 & 521.7389 & Fe 1 & 540.7433 & Fe 2 & 566.2516 & Fe 1 & 606.4620 & Ti 1 & 642.4851 & Ni 1\\
  498.1355 & Ti 2 & 521.9701 & Ti 1 & 541.2784 & Fe 1 & 566.9736 & Si 1 & 606.5482 & Fe 1 & 643.0845 & Fe 1\\
  498.1730 & Ti 1 & 522.0290 & Ni 1 & 541.4070 & Fe 2 & 566.9943 & Ni 1 & 608.1445 & Fe 2 & 643.2676 & Fe 2\\
  498.2499 & Fe 1 & 522.4300 & Ti 1 & 542.0358 & Fe 2 & 568.2199 & Ni 1 & 608.5258 & Fe 1 & 643.9075 & Ca 1\\
  498.3853 & Fe 1 & 522.4540 & Ti 1 & 542.4068 & Fe 1 & 568.4484 & Si 1 & 608.6282 & Ni 1 & 645.2359 & Fe 2\\
  498.4629 & Fe 1 & 522.5526 & Fe 1 & 542.4645 & Ni 1 & 568.9460 & Ti 1 & 609.0226 & Fe 2 & 645.5598 & Ca 1\\
  498.5983 & Fe 1 & 522.6538 & Ti 2 & 542.5249 & Fe 2 & 569.0425 & Si 1 & 609.1171 & Ti 1 & 646.2567 & Ca 1\\
  498.6903 & Fe 1 & 523.4623 & Fe 2 & 542.6286 & Fe 1 & 569.4740 & Cr 1 & 609.6664 & Fe 1 & 646.9192 & Fe 1\\
  499.1268 & Fe 1 & 523.5363 & Fe 1 & 542.9137 & Ti 1 & 570.1104 & Si 1 & 610.0271 & Fe 1 & 649.1566 & Ti 2\\
  499.2785 & Fe 1 & 523.8586 & Fe 2 & 542.9696 & Fe 1 & 570.1544 & Fe 1 & 610.3220 & Fe 2 & 649.4980 & Fe 1\\
  499.7097 & Ti 1 & 524.2491 & Fe 1 & 543.2511 & Fe 2 & 570.3570 & V 1 & 611.9565 & Fe 1 & 649.5741 & Fe 1\\
  499.9503 & Ti 1 & 524.3776 & Fe 1 & 543.2948 & Fe 1 & 570.5464 & Fe 1 & 612.5021 & Si 1 & 649.6466 & Fe 1\\
  500.0730 & Fe 2 & 524.6768 & Cr 2 & 543.4524 & Fe 1 & 570.7049 & Fe 1 & 612.6219 & Fe 1 & 651.8366 & Fe 1\\
  500.1479 & Ca 2 & 524.7565 & Cr 1 & 544.5042 & Fe 1 & 570.8400 & Si 1 & 613.1852 & Si 1 & 655.4191 & Fe 2\\
  500.2792 & Fe 1 & 525.3021 & Fe 1 & 544.6916 & Fe 1 & 571.2131 & Fe 1 & 613.5362 & Fe 1 & 655.6113 & Fe 2\\
  500.3741 & Ni 1 & 525.3462 & Fe 1 & 546.0492 & Fe 2 & 574.8351 & Ni 1 & 613.6615 & Fe 1 & 657.2790 & Fe 1\\
  500.4044 & Fe 1 & 525.6932 & Fe 2 & 546.0873 & Fe 1 & 575.3122 & Fe 1 & 613.6994 & Fe 1 & 659.9120 & Si 2\\
  500.9645 & Ti 1 & 525.7655 & Fe 1 & 546.6396 & Fe 1 & 576.0344 & Fe 1 & 614.5016 & Si 1 & 660.8025 & Fe 1\\
  501.4942 & Fe 1 & 526.0387 & Ca 1 & 546.6987 & Fe 1 & 576.2391 & Fe 2 & 615.1617 & Fe 1 & 661.3759 & Cr 1\\
  501.6161 & Ti 1 & 526.3306 & Fe 1 & 547.2709 & Fe 1 & 577.2146 & Si 1 & 616.1297 & Ca 1 & 666.1075 & Cr 1\\
  502.0026 & Ti 1 & 526.4802 & Fe 2 & 547.3163 & Fe 1 & 577.8453 & Fe 1 & 616.3424 & Ni 1 & 666.3441 & Fe 1\\
  502.3186 & Fe 1 & 526.5148 & Cr 1 & 547.3900 & Fe 1 & 578.0600 & Fe 1 & 616.5360 & Fe 1 & 666.7710 & Fe 1\\
  502.4844 & Ti 1 & 526.5651 & V 1 & 547.4223 & Ti 1 & 578.1751 & Cr 1 & 616.9563 & Ca 1 & 667.7985 & Fe 1\\
  502.8126 & Fe 1 & 526.7269 & Fe 1 & 547.6321 & Fe 2 & 578.3850 & Cr 1 & 617.7255 & Fe 1 & 671.0318 & Fe 1\\
  502.9618 & Fe 1 & 526.8608 & Ti 2 & 547.7712 & Fe 1 & 578.4658 & Fe 1 & 618.0203 & Fe 1 & 672.1848 & Si 1\\
  503.0778 & Fe 1 & 526.9537 & Fe 1 & 548.1243 & Fe 1 & 578.4969 & Cr 1 & 618.6711 & Ni 1 & 673.9520 & Fe 1\\
  503.1914 & Fe 1 & 528.3621 & Fe 1 & 548.1873 & Fe 1 & 579.3073 & Si 1 & 618.7989 & Fe 1 & 674.3107 & V 1\\
  503.6922 & Fe 1 & 528.4425 & Fe 1 & 548.3099 & Fe 1 & 579.3915 & Fe 1 & 619.1200 & Fe 2 & 675.2707 & Fe 1\\
  503.9957 & Ti 1 & 529.5776 & Ti 1 & 548.7145 & Fe 1 & 579.8171 & Fe 1 & 619.5433 & Si 1 & 679.3258 & Fe 1\\
  504.4211 & Fe 1 & 529.8776 & Fe 1 & 549.0148 & Ti 1 & 580.4034 & Fe 1 & 619.9226 & Fe 2 &  & \\
  504.8436 & Fe 1 & 530.0939 & Fe 2 & 549.0714 & Fe 2 & 580.5217 & Ni 1 & 620.4600 & Ni 1 &  & \\
  506.5985 & Ti 1 & 530.2300 & Fe 1 & 549.1832 & Fe 1 & 581.1914 & Fe 1 & 621.9280 & Fe 1 &  & \\
  506.7713 & Cr 1 & 530.4180 & Cr 1 & 550.3895 & Ti 1 & 583.1596 & Ni 1 & 622.3981 & Ni 1 &  & \\
  506.9090 & Ti 2 & 531.0686 & Cr 2 & 550.4088 & Ni 1 & 583.7701 & Fe 1 & 623.0722 & Fe 1 &  & \\
\hline\end{tabular}
\end{table*}

\begin{table*}
\caption{Lines used to calculate chemical abundances of J01020100-7122208.}
\label{tab:regions_chemical}
\begin{tabular}{|r|l|r|l|r|l|r|l|r|l|}
\hline
  \multicolumn{1}{|c|}{Wavelength} &
  \multicolumn{1}{c|}{Element} &
  \multicolumn{1}{c|}{Wavelength} &
  \multicolumn{1}{c|}{Element} &
  \multicolumn{1}{c|}{Wavelength} &
  \multicolumn{1}{c|}{Element} &
  \multicolumn{1}{c|}{Wavelength} &
  \multicolumn{1}{c|}{Element} &
  \multicolumn{1}{c|}{Wavelength} &
  \multicolumn{1}{c|}{Element} \\
\hline
  526.1704 & Ca 1 & 491.8994 & Fe 1 & 557.2842 & Fe 1 & 602.1820 & Mn 1 & 580.4259 & Ti 1\\
  551.2980 & Ca 1 & 492.4301 & Fe 1 & 557.3102 & Fe 1 & 481.1983 & Ni 1 & 586.6451 & Ti 1\\
  559.0114 & Ca 1 & 493.8814 & Fe 1 & 557.6089 & Fe 1 & 496.5167 & Ni 1 & 590.3315 & Ti 1\\
  586.7562 & Ca 1 & 494.5636 & Fe 1 & 558.6756 & Fe 1 & 497.6130 & Ni 1 & 592.2110 & Ti 1\\
  610.2723 & Ca 1 & 494.6387 & Fe 1 & 561.5644 & Fe 1 & 497.6325 & Ni 1 & 596.5828 & Ti 1\\
  615.6023 & Ca 1 & 496.2572 & Fe 1 & 561.8632 & Fe 1 & 539.2331 & Ni 1 & 597.8541 & Ti 1\\
  616.1297 & Ca 1 & 497.0646 & Fe 1 & 563.6696 & Fe 1 & 550.4088 & Ni 1 & 609.1171 & Ti 1\\
  616.3755 & Ca 1 & 497.7648 & Fe 1 & 566.1345 & Fe 1 & 551.0003 & Ni 1 & 612.6216 & Ti 1\\
  616.6439 & Ca 1 & 498.2499 & Fe 1 & 566.2516 & Fe 1 & 558.7858 & Ni 1 & 498.1355 & Ti 2\\
  616.9563 & Ca 1 & 498.3853 & Fe 1 & 570.5464 & Fe 1 & 561.4773 & Ni 1 & 506.9090 & Ti 2\\
  645.5598 & Ca 1 & 498.6223 & Fe 1 & 571.2131 & Fe 1 & 564.1881 & Ni 1 & 533.6786 & Ti 2\\
  647.1662 & Ca 1 & 500.2792 & Fe 1 & 577.5081 & Fe 1 & 564.9699 & Ni 1 & 538.1022 & Ti 2\\
  649.9650 & Ca 1 & 500.4044 & Fe 1 & 577.8453 & Fe 1 & 566.9943 & Ni 1 & 541.8768 & Ti 2\\
  650.8850 & Ca 1 & 500.5712 & Fe 1 & 578.4658 & Fe 1 & 574.8351 & Ni 1 & 480.7521 & V 1\\
  481.3476 & Co 1 & 502.8126 & Fe 1 & 579.3915 & Fe 1 & 580.5217 & Ni 1 & 524.0862 & V 1\\
  481.3972 & Co 1 & 503.1914 & Fe 1 & 581.1914 & Fe 1 & 583.1596 & Ni 1 & 562.4872 & V 1\\
  497.1930 & Co 1 & 504.7126 & Fe 1 & 583.7701 & Fe 1 & 608.6282 & Ni 1 & 562.7633 & V 1\\
  517.6076 & Co 1 & 507.9223 & Fe 1 & 584.9683 & Fe 1 & 618.6711 & Ni 1 & 564.6108 & V 1\\
  523.0208 & Co 1 & 510.4030 & Fe 1 & 585.5076 & Fe 1 & 632.2166 & Ni 1 & 565.7435 & V 1\\
  533.1453 & Co 1 & 521.7389 & Fe 1 & 586.1109 & Fe 1 & 636.6481 & Ni 1 & 566.8361 & V 1\\
  538.1770 & Co 1 & 522.2395 & Fe 1 & 595.2718 & Fe 1 & 637.8247 & Ni 1 & 572.7652 & V 1\\
  548.9662 & Co 1 & 524.2491 & Fe 1 & 603.4035 & Fe 1 & 641.4581 & Ni 1 & 573.7059 & V 1\\
  564.7234 & Co 1 & 524.3776 & Fe 1 & 609.6664 & Fe 1 & 531.8349 & Sc 2 & 608.1441 & V 1\\
  611.6990 & Co 1 & 525.3021 & Fe 1 & 613.6615 & Fe 1 & 533.4240 & Sc 2 & 613.5361 & V 1\\
  495.3717 & Cr 1 & 526.7269 & Fe 1 & 613.6994 & Fe 1 & 568.4202 & Sc 2 & 625.6886 & V 1\\
  506.7713 & Cr 1 & 529.8776 & Fe 1 & 615.1617 & Fe 1 & 624.5637 & Sc 2 & 627.4649 & V 1\\
  512.3460 & Cr 1 & 531.0463 & Fe 1 & 616.5360 & Fe 1 & 660.4601 & Sc 2 & 653.1415 & V 1\\
  524.7565 & Cr 1 & 532.0036 & Fe 1 & 618.7989 & Fe 1 & 564.5613 & Si 1 & 498.2814 & Na 1\\
  526.5148 & Cr 1 & 532.9989 & Fe 1 & 625.2555 & Fe 1 & 566.9736 & Si 1 & 615.4226 & Na 1\\
  527.2000 & Cr 1 & 538.6333 & Fe 1 & 627.1278 & Fe 1 & 568.4484 & Si 1 & 616.0747 & Na 1\\
  530.4180 & Cr 1 & 539.8279 & Fe 1 & 640.0317 & Fe 1 & 613.1852 & Si 1 & 517.2684 & Mg 1\\
  531.2856 & Cr 1 & 541.2784 & Fe 1 & 641.9949 & Fe 1 & 624.3815 & Si 1 & 518.3604 & Mg 1\\
  531.8771 & Cr 1 & 546.6396 & Fe 1 & 662.5022 & Fe 1 & 640.7291 & Si 1 & 669.6023 & Al 1\\
  532.9138 & Cr 1 & 547.2709 & Fe 1 & 666.7710 & Fe 1 & 641.4980 & Si 1 & 669.8673 & Al 1\\
  532.9784 & Cr 1 & 547.3163 & Fe 1 & 667.7985 & Fe 1 & 498.1730 & Ti 1 & 521.8197 & Cu 1\\
  534.0447 & Cr 1 & 547.3900 & Fe 1 & 669.9141 & Fe 1 & 500.9645 & Ti 1 & 488.3682 & Y 2\\
  534.4756 & Cr 1 & 548.7145 & Fe 1 & 679.3258 & Fe 1 & 501.6161 & Ti 1 & 512.3211 & Y 2\\
  538.6968 & Cr 1 & 549.1832 & Fe 1 & 526.4802 & Fe 2 & 522.4540 & Ti 1 & 532.0782 & Y 2\\
  562.8642 & Cr 1 & 552.4250 & Fe 1 & 532.5552 & Fe 2 & 528.8794 & Ti 1 & 572.8886 & Y 2\\
  564.8261 & Cr 1 & 553.6580 & Fe 1 & 541.4070 & Fe 2 & 529.5776 & Ti 1 & 585.3668 & Ba 2\\
  569.4740 & Cr 1 & 553.8516 & Fe 1 & 542.5249 & Fe 2 & 542.9137 & Ti 1 & 614.1713 & Ba 2\\
  578.3065 & Cr 1 & 553.9280 & Fe 1 & 643.2676 & Fe 2 & 550.3895 & Ti 1 & 412.9700 & Eu 2\\
  578.7919 & Cr 1 & 554.6990 & Fe 1 & 482.3520 & Mn 1 & 567.9916 & Ti 1 & 664.5100 & Eu 2\\
  578.8381 & Cr 1 & 554.9949 & Fe 1 & 538.8503 & Mn 1 & 568.9460 & Ti 1 &  & \\
  480.0649 & Fe 1 & 556.9618 & Fe 1 & 542.0351 & Mn 1 & 570.2660 & Ti 1 &  & \\
  481.5230 & Fe 1 & 557.0051 & Fe 1 & 551.6766 & Mn 1 & 571.6450 & Ti 1 &  & \\
\hline\end{tabular}

\end{table*}

\section{Comments on the determination of physical parameters}\label{sec:Hlines}

\subsection{Comparison with observed spectra of control stars}
\begin{figure*}
\centering
\includegraphics[width=5.5cm]{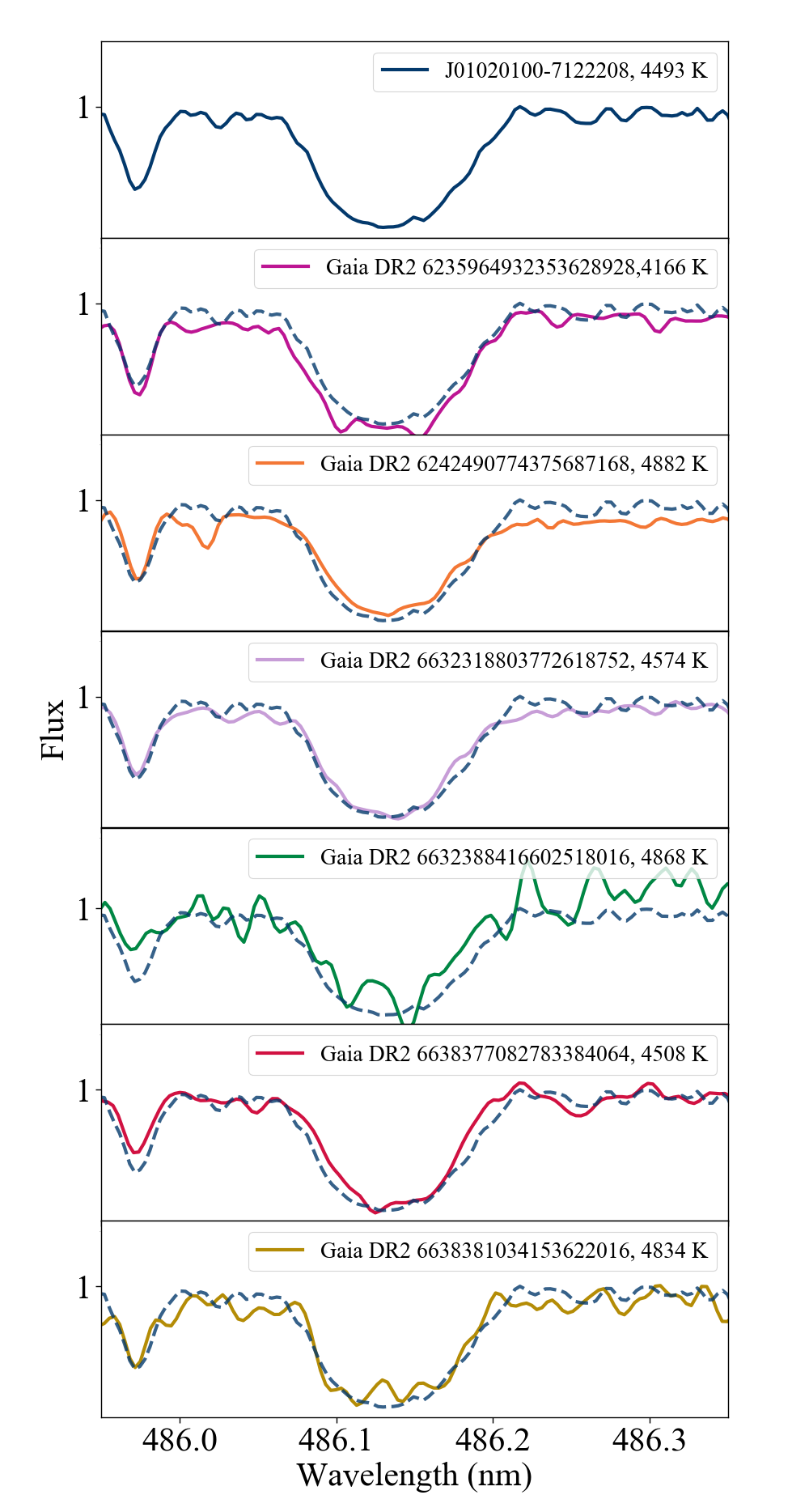}
\includegraphics[width=5.5cm]{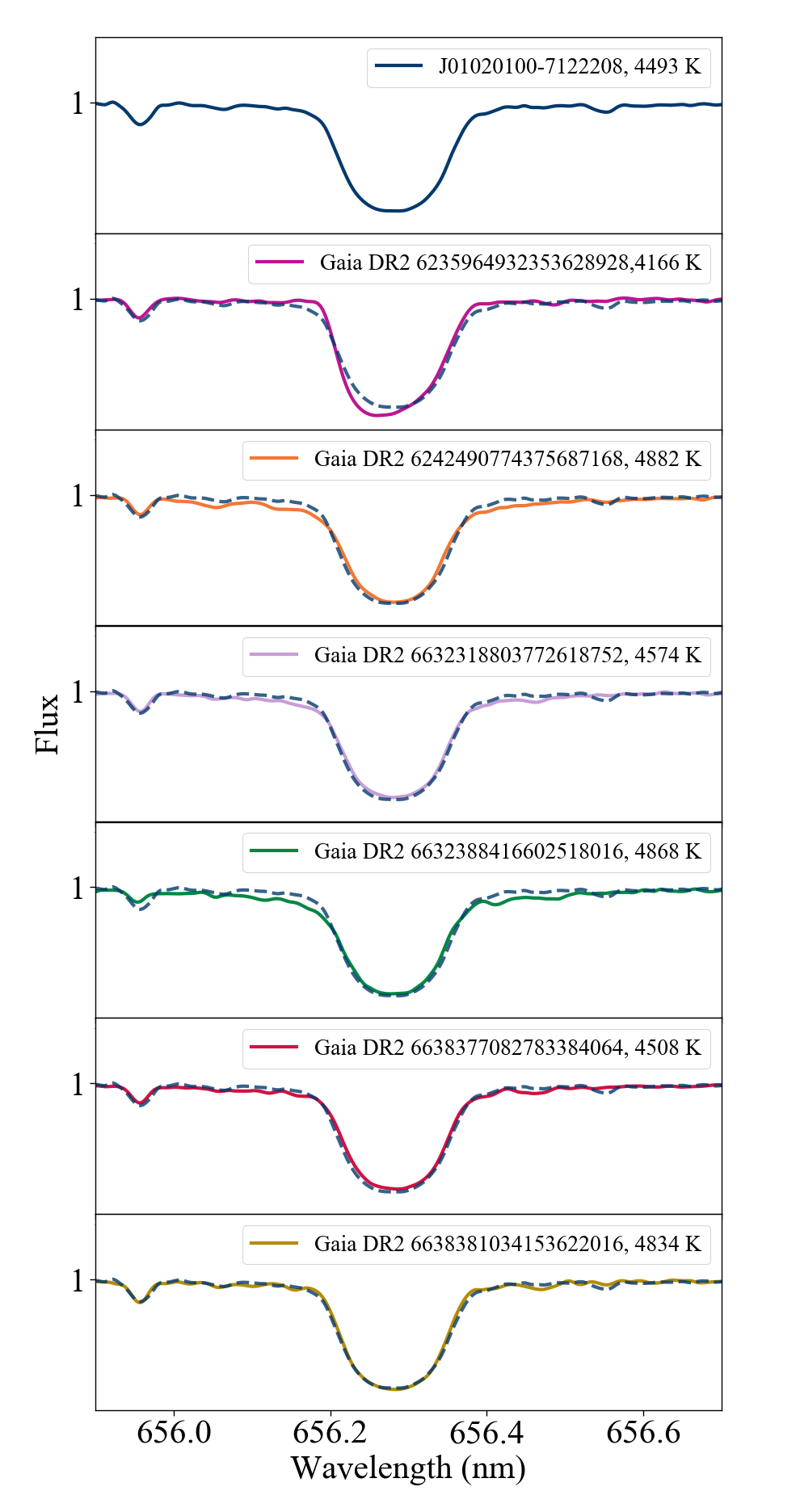}\\
\includegraphics[width=5.5cm]{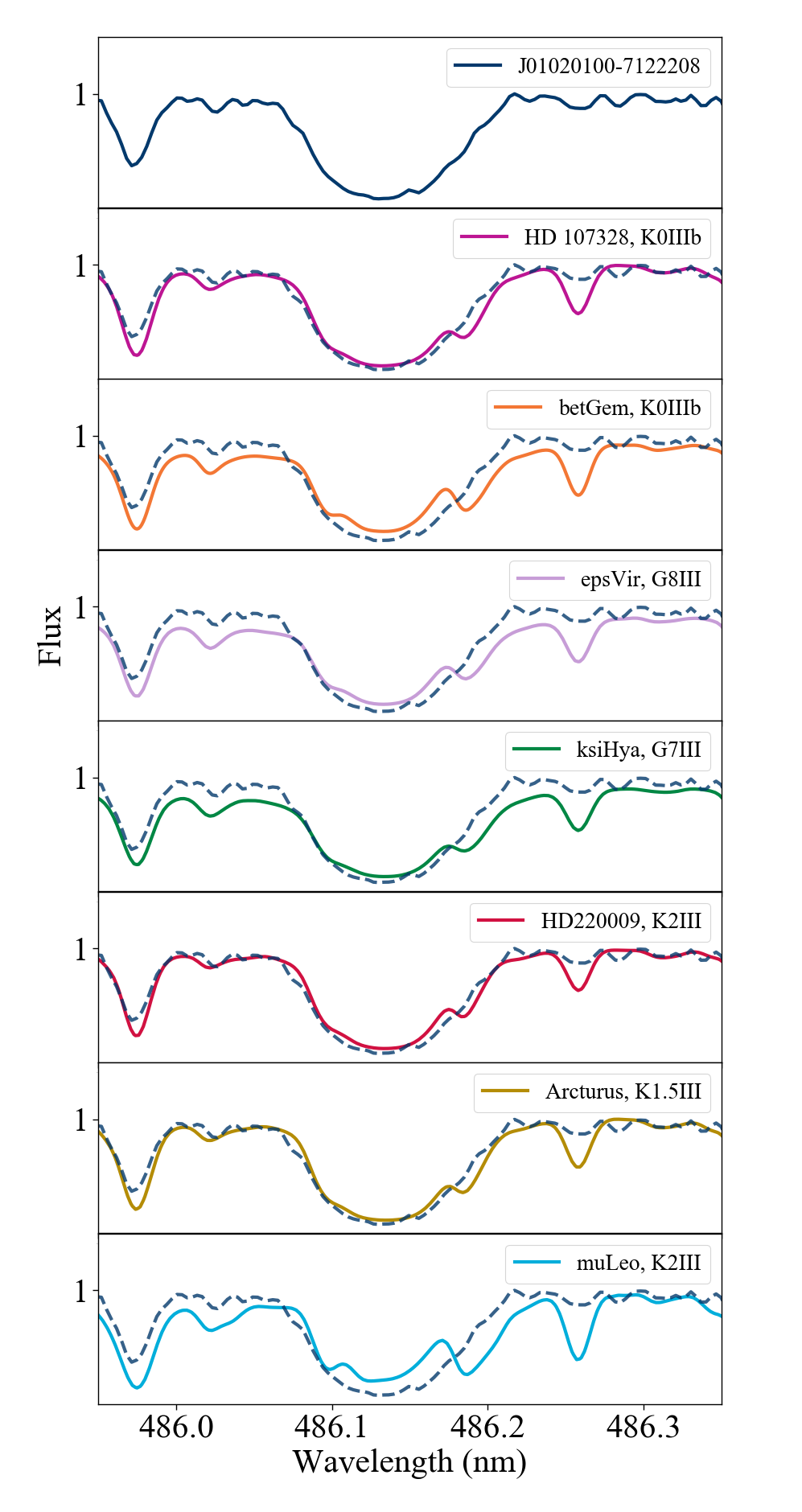}
\includegraphics[width=5.5cm]{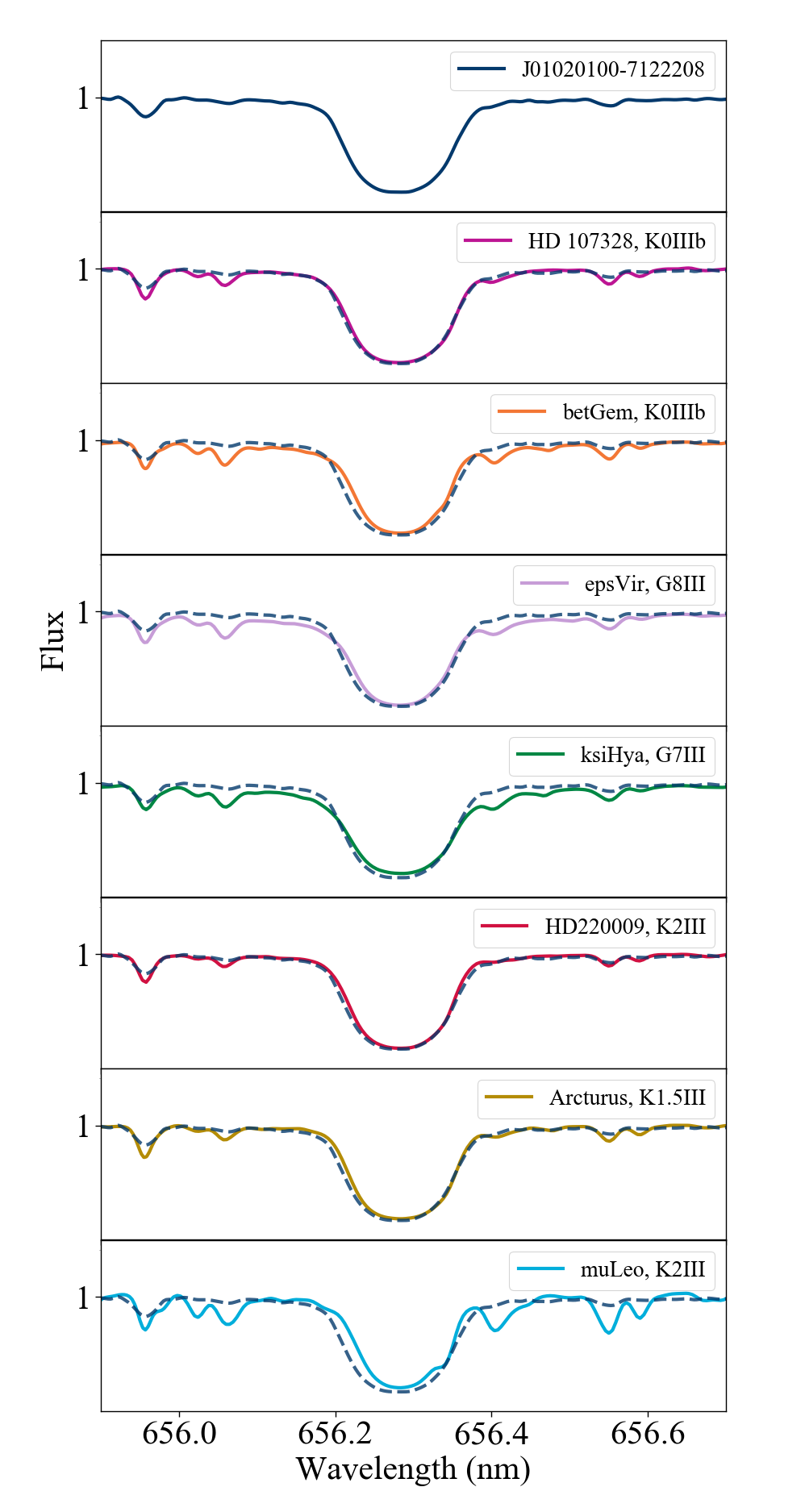}
\caption{Top left: Line profile of H$\beta$ for J01020100-7122208 (first spectrum) and for other giant stars in our control sample. Top right: Line profile of H$\alpha$ for J01020100-7122208 (first spectrum) and for other giant stars in our control sample. Bottom left: Line profile of H$\beta$ for J01020100-7122208 (first spectrum) and for other giant Gaia Benchmark stars. Bottom right: Line profile of H$\alpha$ for J01020100-7122208 (first spectrum) and for other giant Gaia Benchmark stars. In all panels containing control sample stars or GBS, the line profile of J01020100-7122208 is represented as blue dashed lines in order to help with the comparison.}
\label{fig:Hlines_control_GBS}
\end{figure*}

As a sanity check, we visually inspected at the profiles of the H lines, as was done previously in \cite{neugent2018runaway} and \cite{massey2018runaway}. The first verification we did was to compare the regions containing H$\alpha$ and H$\beta$ lines of J01020100-7122208 with the H profiles of giant stars from our control sample, as well as those of the of Gaia Benchmark Stars (GBS) spectral library \citep{blanco2014gaia}. We chose that library because GBS have known spectral type and their parameters are used to validate several current spectroscopic surveys pipelines  \citep{heiter2015benchmark,jofre2014gaia,jofre2015gaia,hawkins2016gaia,jofre2017gaia}. For this purpose, we selected the giant GBS and considered the stellar classification of \cite{heiter2015benchmark}. We did not use H lines located in the bluest regions because the spectrum of J01020100-7122208 is very noisy there and the spectra of GBS do not contain that region. The studied H lines are presented in Figure~\ref{fig:Hlines_control_GBS}. We can see that likely there is a degeneracy in H$\alpha$, since the profile is very similar among all the stars, independently of their $T_\mathrm{eff}$. We can break the degeneracy with the other regions of the high resolution spectrum, because we use many iron lines of different ionisation and excitation states.  We also note that the H profile of J01020100-7122208 agrees well with the profile of K0 stars from the GBS sample, and also with stars with $T_\mathrm{eff}$ of approximately 4500 K in the control sample.

\subsection{Comparison with spectra}

\begin{figure*}
\centering
\includegraphics[width=8cm]{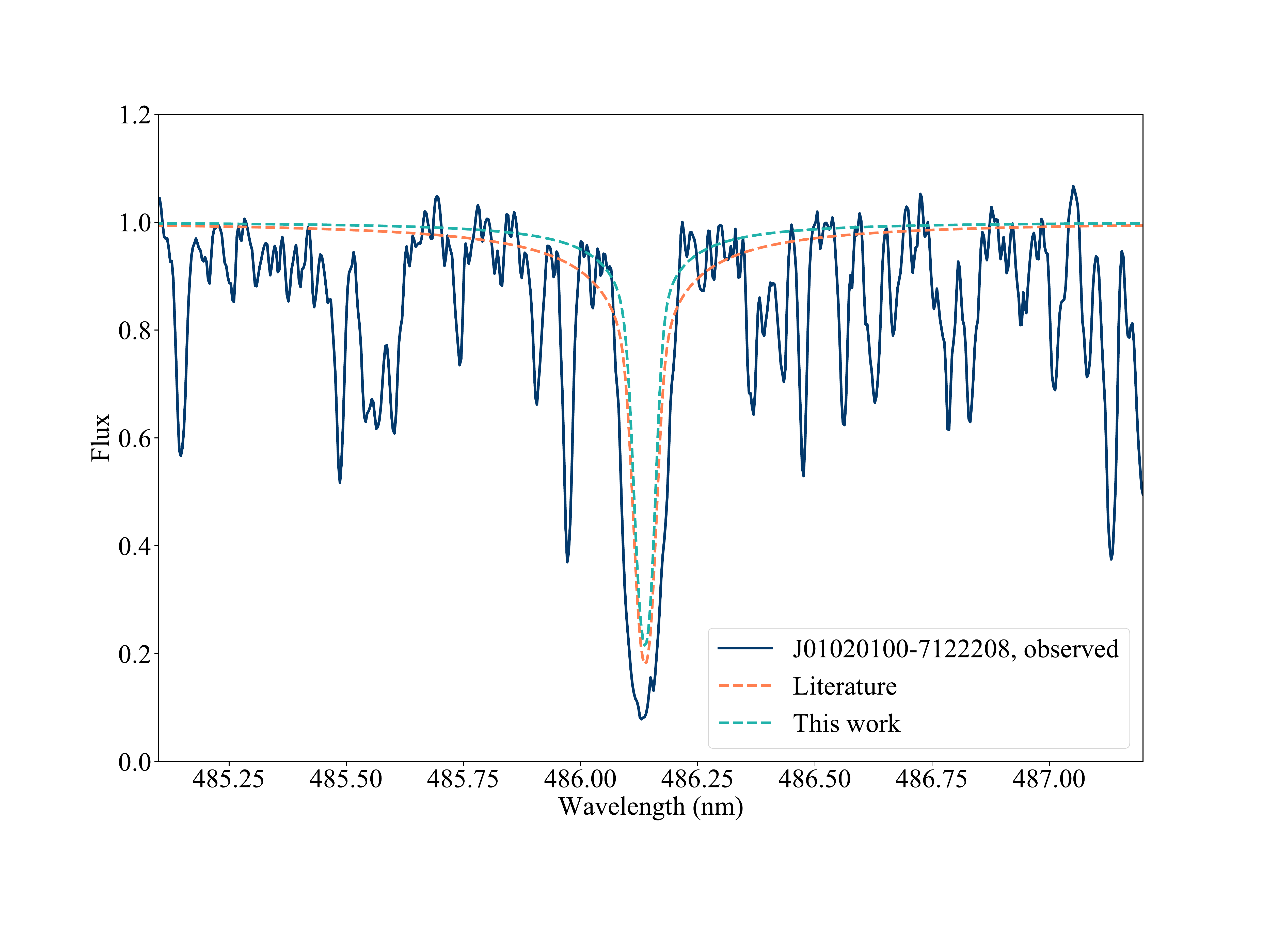}
\includegraphics[width=8cm]{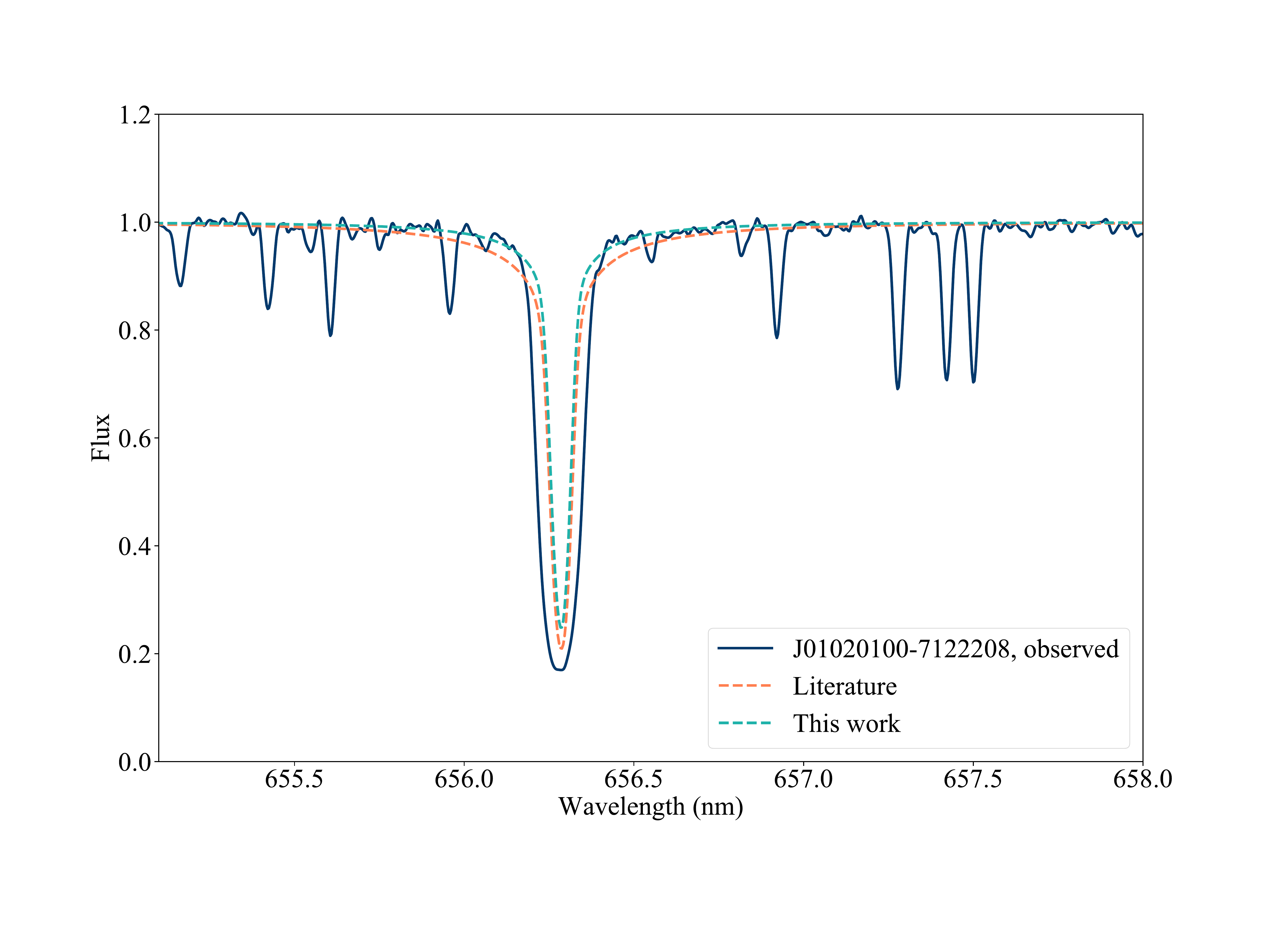}
\caption{Left: H$\beta$ profile of J01020100-7122208. Solid blue line represents the observed spectrum, the dashed green line represents a synthetic spectrum from the grid presented in \protect\cite{amarsi2018effective} with similar stellar parameters of those reported in our work and the dashed orange line represents another synthetic spectrum from the grid of \protect\cite{amarsi2018effective}, but with similar stellar parameters of those presented in \protect\cite{massey2018runaway}.
Right: H$\alpha$ profile of J01020100-7122208. Solid blue line represents the observed spectrum, the dashed green line represents a synthetic spectrum from the grid presented in \protect\cite{amarsi2018effective} with similar stellar parameters of those reported in our work and the dashed orange line represents another synthetic spectrum from the grid of \protect\cite{amarsi2018effective}, but with similar stellar parameters of those presented in \protect\cite{massey2018runaway}.}
\label{fig:Hlines_synthesis}
\end{figure*}

H I Balmer lines of FGK stars are useful when determining the temperature of stars (e.g. \citet{searle1962effective,gehren1981temperature,ruchti2013unveiling,amarsi2018effective}). The wings of these lines are weakly dependent on the surface gravity of the star and the metallicity, being almost only sensitive to the temperature of the gas.  Due to uncertainties in the models and observations, it is challenging to create H I profiles, but by exploring the wings of theses lines, we can obtain a good approximation of $T_\mathrm{eff}$ for a FGK-type star \citep{Jofre2019}.
We used the synthetic grid from \cite{amarsi2018effective}, where the authors created a grid considering 3D and NLTE. The grid contains the regions of H$\alpha$ and H$\beta$.
We chose the spectrum of a stars with $T_\mathrm{eff} = 4500$ K, $\log g = 1.5$ and $\mathrm{[Fe/H]} = -1.25$ to represent the results obtained in our work and a spectrum with $T_\mathrm{eff} = 4800$ K, $\log g = 2.0$ and $\mathrm{[Fe/H]} = -0.5$ to represent those of \cite{massey2018runaway}.
The profiles of H lines are presented in Figure \ref{fig:Hlines_synthesis}. In this figure we observe that despite none of the synthetic spectra reproduce the line accurately (likely due to model limitations which affect the broadening of the lines), the wings of both H$\alpha$ and H$\beta$ are better represented by the stellar parameter values reported in our work.

\bsp
\label{lastpage}
\end{document}